\documentclass[12pt,a4paper]{article}
\usepackage[utf8]{inputenc}
\usepackage[english]{babel}
\usepackage[left=30mm,right=30mm,top=25mm,bottom=35mm
]{geometry}
\usepackage[dvipsnames]{xcolor}
\usepackage{amssymb, amsmath, bbold, mathbbol}
\usepackage{hyperref}
\hypersetup{
    colorlinks=true,
    linkcolor=OrangeRed,
    urlcolor=ForestGreen,
    citecolor=Aquamarine
}
\usepackage{comment}
\numberwithin{equation}{section}

\usepackage{cite}

\makeatletter

\newcommand\frontmatter{%
    \cleardoublepage
  \pagenumbering{roman}}

\newcommand\mainmatter{%
    \cleardoublepage
  \pagenumbering{arabic}}

\newcommand\backmatter{%
  \if@openright
    \cleardoublepage
  \else
    \clearpage
  \fi
   }

\makeatother

\begin{document}

\frontmatter

\begin{titlepage}

\vspace*{15pt}
\begin{center}
{\huge Irrelevant Operators and their\\Holographic Anomalies}

\vspace{35pt}
{\large Matteo Broccoli}

\vspace{30pt}

{\it
Max-Planck-Institut für Gravitationsphysik,\\
Albert-Einstein-Institut, \\
Potsdam-Golm, D-14476, 
Germany.
}

 \vspace{35pt}
April 5, 2022

\vspace{45pt}

{ABSTRACT}
\end{center}

\noindent
Irrelevant operators in a CFT modify the usual Weyl transformation of the metric.
A metric beta-function turns on, which modifies the Weyl anomalies as well.
In this paper, we study the relation between bulk diffeomorphisms and Weyl transformation at the boundary when a massive scalar field, which sources irrelevant operators at the boundary, is coupled to the bulk metric.
Considering the effect of the backreaction generated by the scalar field, we provide a holographic description of the boundary metric beta-function and anomalies.
Our results represent an additional test of the AdS/CFT correspondence.
\noindent 

\end{titlepage}


\pagebreak

\mainmatter

\tableofcontents
\noindent\rule{\textwidth}{1pt}

\section{Introduction}

The holographic computation of the Weyl anomaly \cite{Henningson:1998gx} has been one of the very first non-trivial tests of the AdS/CFT correspondence.
Its success relies on the equivalence between bulk diffeomorphisms and Weyl transformations at the boundary of AdS space, which, since then, has been widely studied.
However, it is not so clear how the equivalence works when the asymptotics of AdS are deformed, namely when the fields present at the boundary induce a strong backreaction onto the gravitational background.

In~\cite{Henningson:1998gx} Einstein's equations with a negative cosmological constant are solved in terms of the boundary metric and it is shown that the on-shell action is divergent at the boundary.
The divergences can be cancelled by the addition of local counterterms, but the regularisation spoils the conformal symmetry of the boundary theory and gives rise to a holographic Weyl anomaly.
Holographic renormalisation has been further developed in~\cite{deHaro:2000vlm}, where scalar fields coupled to gravity in the bulk and their contributions to the Weyl anomaly are also considered.
When scalar fields are added on top of a dynamical background, they induce a backreaction, and the method presented in~\cite{deHaro:2000vlm} is consistent when the bulk scalars are dual to relevant or marginal operators of the boundary CFT (see also~\cite{Witten:1998qj}).
Scalars that are dual to irrelevant operators induce a stronger backreaction onto the gravitational background, and in~\cite{vanRees:hr_irr_op} the method of holographic renormalisation has been extended to account for such backreaction, when the irrelevant operators are of non-integer conformal dimension, so that no logarithmic divergence appears in the renormalised action (i.e.~there is no anomaly in the CFT).
Logarithmic divergences are considered in~\cite{vanRees:CS_eq-w_anomalies}, where the conformal anomaly in the three-point function of irrelevant operators is computed in the CFT and derived from holography.
However, in this case no backreaction has been considered and the gravitational background has been taken to be unperturbed by the presence of the scalar fields.
Here, we want to study the effect of coupling integer-dimensional scalars that source irrelevant operators on the boundary theory to a dynamical background, thus including the effect of the backreaction.

CFTs with integer-dimensional irrelevant operators have peculiar properties~\cite{Schwimmer:2019efk}.
In order to have a solution of the Wess-Zumino consistency condition for the Weyl anomaly, the presence of these operators requires a modification of the usual Weyl transformation.
A metric beta-function, which depends on the sources of the irrelevant operators, has to be introduced in the Weyl transformation of the metric.\footnote{In~\cite{Schwimmer:2019efk} the footprint of the metric beta-function is found in the flat space three-point function involving the stress tensor and two irrelevant scalar operators. A thoroughly analysis of this and other CFT three-point correlators is performed in~\cite{Bzowski:2018fql}, where anomalies and beta-functions are also identified. Explicit results for irrelevant scalar operators in $d=4$ are not presented there, although the presence of beta-functions is pointed out.}
As a consequence, the Weyl anomaly is deformed by the metric beta-function, i.e.~the solution of the Wess-Zumino consistency condition is different in the presence of the beta-function.
The geometry, however, is not subject to an RG flow, since correlation functions of irrelevant operators in~\cite{Schwimmer:2019efk} are computed in the undeformed CFT.\footnote{Metric beta-functions in the framework of holographic RG flows are also discussed e.g.~in~\cite{Verlinde:1999xm,deBoer:1999tgo,Jackson:2013eqa,Kiritsis:2014kua,Nakayama:2014cca}.}

Here we will present a holographic description of the four-dimensional CFT studied in~\cite{Schwimmer:2019efk}.
Therefore, we will also generalise the analysis of~\cite{vanRees:CS_eq-w_anomalies} to include the case in which the bulk scalar fields are coupled to a dynamical background.
However, to make contact with~\cite{Schwimmer:2019efk}, we will be interested in describing a boundary theory that is not deformed by the irrelevant operators.
To do so, the tool that we find most convenient to use is that of Penrose-Brown-Henneaux (PBH) transformations~\cite{Imbimbo:1999bj}.
These are a particular class of bulk diffeomorphisms that reduces to Weyl transformation on the boundary. 
They consist of a general transformation rule for the bulk metric, and, as such, they do not require solving any equation of motion.
An action evaluated on a metric that is a solution of the PBH transformation allows to study the Weyl anomaly of the boundary theory.
Thus, we will extend the analysis of~\cite{Imbimbo:1999bj} to include massive scalar fields in the bulk and provide a holographic description of the results obtained in~\cite{Schwimmer:2019efk}.
As we will show, the condition that the irrelevant operators at the boundary do not deform the CFT requires that we are off-shell in the bulk, thus making the PBH transformation an ideal framework for the present analysis.

Irrelevant deformations of CFTs have received attention in particular in the form of $T \bar T$ deformations.
It is proposed that a two-dimensional $T \bar T$ deformed CFT is dual to a three-dimensional AdS space with a sharp cutoff~\cite{McGough:2016lol}.
This conjecture has then been further analysed and extended to higher dimensions~\cite{Guica:2019nzm,Taylor:2018xcy,Hartman:2018tkw}.
Although we are not considering a $T\bar T$ deformation, the scalar field that we will add in the bulk is sourcing an integer-dimensional irrelevant operator on the boundary.
As we will see, this addition will still have the effect of moving the boundary into the bulk, but, since we are interested in describing an undeformed boundary theory, we will have to move the cutoff back to the AdS boundary where the undeformed CFT lives.
In so doing, the solution of the PBH transformations will go off-shell, in the sense that they won't match anymore the solution of the equations of motion of a scalar field coupled to a dynamical background.
Nonetheless, once this is done we will precisely recover the physics described in~\cite{Schwimmer:2019efk}.
We will thus see that the Weyl transformation of the metric is no longer the usual one, but it acquires a beta-function, and we will be able to study the modified anomaly in our holographic set-up.
In the end, this analysis will sharpen the understanding of the AdS/CFT duality in the presence of a backreaction in the bulk theory.

The paper is organized as follows.
In sec.~\ref{sec:PBH} we review the pure gravity formulation of the PBH transformations and how the holographic Weyl anomaly is obtained.
In sec.~\ref{sec:pbhwscalar} we extend the PBH transformations to describe a  scalar field coupled to gravity and choose the scalar field to have integer dimension $\Delta= d+1$.
We show that the transformation of the boundary metric is no longer the usual Weyl transformation and we derive the modified Weyl anomaly of the four-dimensional boundary theory.
These are the main results of the present paper.
We then conclude and discuss possible extensions of our analysis.
We report explicit results that are needed to derive the modified anomaly and its complete expression in the appendices and in the ancillary Mathematica notebook.


\section{Pure gravity}\label{sec:PBH}
We introduce now the PBH transformations for the metric, mainly reviewing~\cite{Imbimbo:1999bj} (see also~\cite{Schwimmer:2000cu,Schwimmer:2003eq,Schwimmer:2008yh,Schwimmer:2008zz}) to set our notation.

\subsection{PBH transformations}
Consider an asymptotically AdS space in $(d+1)$ dimensions with coordinates $(z,x^i)$ such that the bulk metric has the Fefferman-Graham (FG) form\footnote{We choose an AdS space with radius $\ell = 1$. Greek letters are used for $(d+1)$-dimensional bulk indices, while Latin letters for $d$-dimensional boundary indices. Our convention on the curvature is $[\nabla_\mu , \nabla_\nu] V_\rho= R_{\mu\nu\rho}{}{}{}^\sigma V_\sigma$, with $R_{\mu\nu} = R_{\mu\rho\nu}{}{}{}^{\rho}$.}
\begin{equation}\label{fg}
    ds^2 = G_{\mu\nu} dX^\mu dX^\nu = \frac{dz^2}{4z^2} + \frac{1}{z}  g_{ij}(z,x) dx^i dx^j \, ,
\end{equation}
where the boundary is at $z=0$, with $g_{ij}(z=0,x)=g^{(0)}_{ij}(x)$ being the boundary metric.
In order to study the Weyl anomaly of the dual CFT, we first study the behaviour of $G_{\mu\nu}$ under diffeomorphisms.
Under a general coordinate transformation
\begin{equation}\label{diffeo}
    X^\mu = X'^\mu + \xi^\mu (X') \, ,
\end{equation}
$G_{\mu\nu}$ transforms as
\begin{equation}\label{Gtransf}
    \delta G_{\mu\nu} = G_{\mu\rho} \partial_\nu \xi^\rho + G_{\nu\rho} \partial_\mu \xi^\rho + \xi^\rho \partial_\rho G_{\mu\nu} \, ,
\end{equation}
with $\delta G_{\mu\nu} = \mathcal{L}_\xi G_{\mu\nu} = G'_{\mu\nu}(z,x) - G_{\mu\nu}(z,x)$
and we require that $\xi^\mu$ is such that~\eqref{fg} is form invariant under~\eqref{diffeo}, i.e.~$\mathcal{L}_\xi G_{zz} = 0 = \mathcal{L}_\xi G_{zi}$.
The solution is given by
\begin{equation}\label{a}
    \xi^z = -2 z \sigma(x) \, , \quad \xi^i = a^i(z,x) = \frac12 \partial_j \sigma(x) \int_0^z dz' \,   g^{ij}(z',x) \, ,
\end{equation}
where $\sigma(x)$ is an arbitrary function, the $a^i$ are infinitesimal and we will work to order $\mathcal{O}(\sigma, a^i)$.
The lower end of the integration is chosen so that there are no residual diffeomorphisms at the boundary $z=0$.
In other words, the boundary condition $a^i(z=0,x) =0$ holds.
From form invariance of $G_{ij}$ it follows that
\begin{equation}\label{g}
    \delta g_{ij} = 2\sigma (1-z\partial_z) g_{ij} + \nabla_i a_j + \nabla_j a_i \, ,
\end{equation}
where indices are lowered with (and derivatives are covariant w.r.t.) $g_{ij}$.
Eqs.~\eqref{a}, \eqref{g} define the PBH transformations.

We show now that this particular class of bulk diffeomorphisms reduces to a Weyl transformation at the boundary.
Indeed, the commutator of two diffeomorphisms~\eqref{Gtransf} is again a diffeomorphism
\begin{equation}\label{commdiffeo}
    [\delta_2,\delta_1] G_{\mu\nu} = G_{\mu\rho} \partial_\nu \hat\xi^\rho + G_{\nu\rho} \partial_\mu \hat\xi^\rho + \hat\xi^\rho \partial_\rho G_{\mu\nu} \, ,
\end{equation}
where $\hat\xi^\rho$ is defined as
\begin{equation}\label{xihat}
    \hat\xi^\rho = \xi^\sigma_1 \partial_\sigma \xi^\rho_2 - \xi^\sigma_2 \partial_\sigma \xi^\rho_1 + \delta_2 \xi^\rho_1 - \delta_1 \xi^\rho_2
\end{equation}
and the last two terms are non-vanishing if we allow $\xi^\mu$ to be field dependent.
If the diffeomorphism is a PBH, then it is possible to derive the PBH group property $\hat\xi^\mu = 0$ \cite{Schwimmer:2008yh,Fiorucci:2020xto}, so that $[\delta_2,\delta_1] G_{\mu\nu} = 0$.
Since the PBH transformations do not act on coordinates, it follows that \mbox{$[\delta_2,\delta_1] g_{ij} = 0$} and the PBH transformations reduce to a Weyl transformation on the boundary metric.
Indeed, from~\eqref{g} at $z=0$ it follows that
\begin{equation}
    \delta g^{(0)}_{ij} = 2 \sigma g^{(0)}_{ij} \, .
\end{equation}

Now we assume the following power series expansions about the boundary\footnote{If $d$ is an even integer, the expansion of the metric contains also logarithmic terms.
Here we work in generic dimensions, and we do not need to include such terms.}
\begin{align}
    a^i(z,x) &= \sum_{n=1}^\infty a^i_{(n)}(x) z^n \label{aansatz}\\
    g_{ij}(z,x) &= \sum _{n=0}^\infty g^{(n)}_{ij}(x) z^n \, . \label{gansatz}
\end{align}
Using the PBH equations, we can determine the coefficient in the expansion of the metric in terms of covariant tensors built from the boundary metric as follows.
First, we compute the $a_{(n)}$ in terms of the $g_{(n)}$, and for the first few terms we find
\begin{align}
    a^i_{(1)} &= \frac12 g_{(0)}^{ij}\partial_j \sigma \label{a1}\\
    a^i_{(2)} &= -\frac14 g_{(1)}^{ij}\partial_j \sigma \label{a2} \, .
\end{align}
Indices are now lowered (raised) with the (inverse of) $g^{(0)}_{ij}$; curvatures and covariant derivatives will be w.r.t.~$g^{(0)}_{ij}$.
Then, combining the expansions and the $a_{(n)}$ into~\eqref{g} we find the variation of $g_{(n)}$ as
\begin{align}
    \delta g^{(0)}_{ij} =& \, 2\sigma g^{(0)}_{ij} \label{dg0}\\
    \vdots & \nonumber\\
    \delta g^{(n)}_{ij} =& \, 2\sigma (1-n) g^{(n)}_{ij} + \sum_{m=1}^{n} \left( g^{(n-m)}_{ik} \partial_j a^k_{(m)} + g^{(n-m)}_{jk} \partial_i a^k_{(m)} + a^k_{(m)} \partial_ k g^{(n-m)}_{ij} \right)
\end{align}
and we see that in general $g_{(n)}$ contains $2n$ derivatives.
Thus, to compute it we make the most general Ansatz for a symmetric tensor with two indices and with $2n$ derivatives, take its variation according to~\eqref{dg0} and impose that it satisfies the PBH equation.
In this way, for example for the first two terms we find
\begin{align}\label{g1}
    g^{(1)}_{ij} &= - \frac{1}{d-2} \bigg( \overset{(0)}{R}_{ij} -\frac{1}{2(d-1)}\overset{(0)}{R} g^{(0)}_{ij} \bigg) \displaybreak[0] \\
    g^{(2)}_{ij} &= \frac{1}{d-4} \Bigg( \frac{1}{8(d-1)}\overset{(0)}{\nabla}_i \overset{(0)}{\nabla}_j \overset{(0)}{R} - \frac{1}{4(d-2)} \overset{(0)}{\Box} \overset{(0)}{R}_{ij} + \frac{1}{8(d-2)(d-1)}g_{(0)ij} \overset{(0)}{\Box}\overset{(0)}{ R}  \nonumber\\
    &\phantom{=} - \frac{1}{2(d-2)} \overset{(0)}{R}{}^{kl}\overset{(0)}{R}_{ikjl} + \frac{d-4}{2(d-2)^2}\overset{(0)}{R}_{ik}\overset{(0)}{R}_j{}^k + \frac{1}{(d-2)^2(d-1)} \overset{(0)}{R}_{ij} \overset{(0)}{R}  \nonumber \\
    & \phantom{=} + \frac{1}{4(d-2)^2} \overset{(0)}{R}_{kl}\overset{(0)}{R}{}^{kl} g^{(0)}_{ij} - \frac{3d}{16(d-2)^2(d-1)^2} g^{(0)}_{ij} \overset{(0)}{R}{}^{2} \Bigg) \nonumber \\
    &\phantom{=} + c_1 \, \overset{(0)}{C}{}^{2} g^{(0)}_{ij} + c_2 \, \overset{(0)}{C}_{iklm}\overset{(0)}{C}_j{}^{klm} \label{g2}
\end{align}
where $C_{ijkl}$ is the Weyl tensor.
Starting from $g_{(2)}$, the solutions will have free coefficients that are not fixed by the PBH equation.
On the other hand, if one solves Einstein's equations of motion for the metric, $g_{(2)}$ is completely determined in generic dimensions.
The free coefficients in the PBH solutions are thus fixed on-shell given an action.


\subsection{Effective boundary action and Weyl anomalies}\label{sec:PBHanomaly}
Consider now an action
\begin{equation} \label{puregravity}
    S = \int_M d^{d+1} X \sqrt{G} f(R(G)) \, ,
\end{equation}
where $f$ is a local function of the curvature and its covariant derivatives and we require that $f(R)$ is such that the equations of motion are solved by asymptotically $\text{AdS}_{d+1}$ in order to have a CFT at the boundary.
Under a bulk diffeomorphism, the action $S$ is invariant up to a boundary term
\begin{equation}\label{deltaS}
    \delta S = \int_M d^{d+1} X \, \partial_\mu \left( \xi^\mu \mathcal{L} \right) \, , \quad \mathcal{L} = \sqrt{G} f(R(G))
\end{equation}
from which we read the transformation $\delta\mathcal{L} = \partial_\mu \left( \xi^\mu \mathcal{L} \right)$. If the diffeomorphism is a PBH, then one shows that $[\delta_2, \delta_1]\mathcal{L}=0$ upon using the PBH group property \cite{Schwimmer:2008yh}.
In FG coordinates, the metric expansion~\eqref{gansatz} induces a power series expansion for $\mathcal{L}$ as well
\begin{equation}
    \mathcal{L} = \sqrt{g_{(0)}} z^{-d/2-1} \mathcal{L}_g \, , \quad  \mathcal{L}_g = \sum_{n=0}^\infty \mathcal{L}^{(n)}_g (x) z^n \, .
\end{equation}
Then, by virtue of $[\delta_2, \delta_1]\mathcal{L}=0$, it is possible to show that $\mathcal{L}_g$ satisfies a Wess-Zumino condition
\begin{equation}
    \int_{\partial M} d^d x \sqrt{g_{(0)}} \left( \sigma_2(x) \delta_1 \mathcal{L}_g - \sigma_1(x) \delta_2 \mathcal{L}_g \right) = 0 \, ,
\end{equation}
which means that $\mathcal{L}_g$ is a candidate for the anomaly of the boundary CFT.
To make the connection precise, from~\eqref{deltaS} we have
\begin{equation}\label{deltaSbdy}
    \delta S = \int_{\partial M} d^d x \, \xi^z \mathcal{L} |_{z=0} = -2 \int_{\partial M} d^d x \, z \, \sigma \, \mathcal{L} |_{z=0}
\end{equation}
restricting the diffeomorphism to a PBH.
Using the holographic dictionary, we interpret the variation of the bulk action as the variation of the generating functional of the CFT correlators, so that the finite piece in~\eqref{deltaSbdy} gives the holographic Weyl anomaly.\footnote{Divergent terms can be cancelled with the addition of counterterms to the bulk action, as first shown for the Einstein-Hilbert action in~\cite{Henningson:1998gx} and then further elaborated upon and proven for a generic action in~\cite{deHaro:2000vlm,Papadimitriou:2004ap,Papadimitriou:2005ii,Andrade:2006pg}.}
Thus, $\mathcal{L}^{(n)}_g(x)$ is the trace anomaly of the $d=2n$ dimensional CFT.

Consider now as an example the following action
\begin{equation}\label{EHwriem}
    2\kappa^2 f(R(G)) = \Lambda - R(G) + \gamma \left(R_{\mu\nu\rho\sigma}R^{\mu\nu\rho\sigma}\right) \left( G \right) \, , \quad \Lambda = -d (d-1) -2d (d-3)\gamma \, ,
\end{equation}
where $\gamma$ is a dimensionless parameter and $\Lambda$ is such that $\text{AdS}_{d+1}$ with radius $\ell =1$ is a solution of the equations of motion (for simplicity in the following we will take $2\kappa^2 = 16\pi G^{(d+1)}_N =1$).
By writing the action in FG coordinates we find (a prime denotes a derivative w.r.t.~$z$)
\begin{align}
    \mathcal{L}_g &= d (1 +4 \gamma ) + \frac12 z (1+4\gamma) \left[ 2(1 - d) g^{ij}g'_{ij} - R(g) \right]  + \frac12 z^2 \big[ 4(1 + 4 \gamma ) g^{ij} g''_{ij}\nonumber\\
    &\phantom{=} - (3 - 4 (d-5) \gamma ) g^{ik}g^{jl}g'_{ij} g'_{kl} + (1 + 8 \gamma ) \left( g^{ij} g'_{ij} \right)^2  + \gamma (R_{ijkl} R^{ijkl})(g) \nonumber\\
    &\phantom{=} + 8 \gamma g'_{ij} R^{ij}(g) \big] + \gamma z^3 \big[ 4 g^{im}g^{jn}g^{kl} g'_{ik} g'_{mn} g'_{jl}  - 4 g^{ij} g^{kl} g^{mn} g'_{ij} g'_{km} g'_{ln} \nonumber\\
    &\phantom{=} -2  g'_{ij} g'_{kl} R^{ikjl}(g)  - 4 g^{im}g^{jn}\nabla_{j}g'_{ik}\nabla^{k}g'_{mn} + 4 g^{im}g^{jn} \nabla_{k}g'_{ij} \nabla^{k}g'_{mn} \big]  \nonumber\\
    &\phantom{=} + \gamma z^4 \big[ g^{im}g^{jn}g^{kp}g^{lq} g'_{ik} g'_{mn} g'_{jl} g'_{pq} + (g^{ik}g^{jl}g'_{ij} g'_{kl})^2 + 8 g^{ik} g^{jl} g''_{ij} g''_{kl}  \nonumber\\
    &\phantom{=} - 8 g^{im}g^{jn}g^{kl} g'_{ik} g'_{mn} g''_{jl} \big] \label{Og}
\end{align}
and expanding the metric according to~\eqref{gansatz}, we identify for instance the terms
\begingroup
\allowdisplaybreaks[1]
\begin{align}
    \mathcal{L}^{(1)}_g &= - \frac{1}{2} (1+4\gamma) \overset{(0)}{R} + \frac12 \left(2 - d \right)(1+4\gamma) g^{(1)i}{}_{i} \label{Og1}\\
    \mathcal{L}^{(2)}_g &= \frac{1}{2} (1+12\gamma) \overset{(0)}{R}{}^{ij} g^{(1)}{}_{ij}  - \frac{1}{4} (1+4\gamma) \overset{(0)}{R} g^{(1)i}{}_{i} + \frac32 (4 -  d) (1+4\gamma) g^{(2)i}{}_{i}  \nonumber\\*
    &\phantom{=} + \left(\frac18 (8  - 3 d) +\frac32 \gamma (4-d) \right) g^{(1)i}{}_{i} g^{(1)j}{}_{j} - \frac{1}{2}(1+4\gamma) \overset{(0)}{\nabla}_{j}\overset{(0)}{\nabla}_{i}g^{(1)ij}   \nonumber\\*
    &\phantom{=} + \left(\frac14 ( 3 d - 10 ) +\gamma(5d-14) \right) g^{(1)}{}_{ij} g^{(1)ij}  + \frac{1}{2}(1+4\gamma) \overset{(0)}{\Box}g^{(1)i}{}_{i} + \frac12 \gamma \overset{(0)}{R}_{ijkl}\overset{(0)}{R}{}^{ijkl} \label{Og2}
\end{align}
\endgroup
and on the PBH solutions~\eqref{g1},~\eqref{g2} we find the trace anomaly in $d=2,4$ respectively:
\begin{align}
    \mathcal{L}^{(1)}_g &= -\frac12 (1+4\gamma) \overset{(0)}{R} = -\frac12 a E_2\\
    \mathcal{L}^{(2)}_g &= -\frac18 (1+12\gamma) \overset{(0)}{R}{}_{ij} \overset{(0)}{R}{}^{ij} + \frac{1}{24} (1+8\gamma) \overset{(0)}{R}{}^{2} + \frac12 \gamma \overset{(0)}{R}_{ijkl}\overset{(0)}{R}{}^{ijkl} = -\frac{1}{16} \left( c \, C^2 - a \, E_4 \right)
\end{align}
with $c=1-4\gamma$ and $a=1+4\gamma$.
$E_{2n}$ is the Euler density in $d=2n$, and explicitly $E_4 = {R}{}_{ijkl}{R}{}^{ijkl} -4 {R}{}_{ij} {R}{}^{ij} + {R}{}^2$ while $C^2 = {R}{}_{ijkl}{R}{}^{ijkl} -2 {R}{}_{ij} {R}{}^{ij} + \frac13 {R}{}^2$ with curvature w.r.t.~the boundary metric.
Following the classification of~\cite{Deser:1993yx}, we notice that in $d=2$ the anomaly is entirely type A, while in $d=4$ there is also a type B.
In particular, for $\gamma =0$ we have $a=c$, while in the presence of the quadratic term in the curvature in~\eqref{EHwriem} then $a-c\neq 0$.
Since $R^2$ and $R_{\mu\nu}R^{\mu\nu}$ terms in the action would also change the values of $a$ and $c$, but not their difference, \eqref{EHwriem} is the minimal bulk action which allows to distinguish between type A and B anomalies in the pure gravity case.
Having $a\neq c$ will be useful later when we include scalar fields in the bulk.
As a final comment, notice that $g^{(n)}_{ij}$ does not contribute to $\mathcal{L}^{(n)}_g$ in $d=2n$ \cite{Schwimmer:2003eq,Schwimmer:2008yh,Miao:2013nfa}.


\section{Adding a massive scalar field}\label{sec:pbhwscalar}
Now we add a massive scalar field $\Phi$ in the bulk and couple it to the metric.
We want to extend the PBH transformations discussed in sec.~\ref{sec:PBH} to describe this system.\footnote{See also~\cite{Bianchi:2001de,Bianchi:2001kw,Erdmenger:2001ja} for the PBH transformation of scalar fields.}

\subsection{Modified PBH transformations}
From the standard holographic dictionary it is known that a bulk scalar field of mass $m$ is dual to a scalar operator on the boundary theory with dimension $\Delta$, related to the mass by $m^2 = \Delta(\Delta-d)$.
Close to the boundary, we consider the following expansion 
\begin{equation}\label{phiansatz}
    \Phi(z,x) = z^{(d-\Delta)/2} \phi(z,x) \, , \quad \phi(z,x) = \sum _{n=0}^\infty \phi_{(n)}(x) z^n 
\end{equation}
with $\phi_{(0)}$ being the source of the boundary operator.\footnote{As for the metric expansion, there are logarithmic terms in the expansion of $\phi(z,x)$ for even integer dimension $d$.
We assume we do not need to include them in the present discussion.}
Requiring that the bulk scalar is indeed a scalar under diffeomorphisms,
\begin{equation}
    \Phi'(z',x') = \Phi(z,x) \, ,
\end{equation}
and choosing the diffeomorphism to be a PBH~\eqref{a}, we obtain the PBH transformation for the field $\phi$ as
\begin{equation}\label{phi}
    \delta \phi = -2\sigma \bigg( \frac{d-\Delta}{2} +z\partial_z \bigg) \phi + a^i \partial_i \phi \, .
\end{equation}
With the expansion in~\eqref{phiansatz}, we get
\begin{equation}
    \delta \phi_{(n)} = -\sigma (d + 2n -\Delta ) + \sum_{m=0}^{n-1} \left( a^i_{(n-m)} \partial_i \phi_{(m)} \right) \, ,
\end{equation}
that to lowest order yields
\begin{equation}\label{dphi0}
    \delta \phi_{(0)} =  \sigma \left( \Delta - d \right) \phi_{(0)} \, ,
\end{equation}
namely the correct transformation of the source of a dimension $\Delta$ operator under Weyl transformation.
Eventually, we want to make contact with the four-dimensional CFT analysed in~\cite{Schwimmer:2019efk}.
Thus, we choose $\Delta= d+1$, and therefore the scalar field is sourcing an irrelevant operator in the CFT.

When we couple the bulk scalar to gravity, the dynamical background will backreact and the metric in FG form will be as follows
\begin{equation}\label{modfg}
    ds^2  = \frac{dz^2}{4z^2} + \frac{1}{z} \left( g_{ij}(z,x) + h_{ij}(z,x) \right) dx^i dx^j \, ,
\end{equation}
where $h_{ij}$ is the backreaction, which depends explicitly on $g_{ij}(z,x)$ and $\Phi(z,x)$.
We are essentially allowing for perturbations of the metric $g_{ij}$ due to the presence of the scalar field $\Phi$, and the metric $g_{ij}$ is then treated as a background, unperturbed, metric.
To first non-trivial order, the backreaction is quadratic in the scalar field, and from now on we will work to order $\mathcal{O}(\sigma,\phi^2)$.
We impose a boundary condition for the backreaction (following \cite{vanRees:hr_irr_op}), namely that the backreaction does not change the boundary metric.
In other words, $g^{(0)}_{ij}$ is still the boundary metric even in the presence of the backreaction.
We will see the effect of this boundary condition later.

We now derive the modifications of the PBH transformations due to the presence of the backreaction by studying the behaviour of~\eqref{modfg} under diffeomorphisms.
The FG form of the bulk metric in~\eqref{modfg} is invariant under the transformation in~\eqref{diffeo} for 
\begin{equation}\label{modxi}
    \xi^z = -2 z \sigma(x) \, , \quad \xi^i = a^i(z,x) + b^i(z,x) = \frac12 \partial_j \sigma(x) \int_\epsilon^z dz' \left( g^{ij}(z',x) - h^{ij}(z',x) \right) \, ,
\end{equation}
where $b^i$ contains the scalar field corrections brought about by the backreaction and is therefore of order~$\mathcal{O}(\sigma,\phi^2)$, while $a^i$ is still of order~$\mathcal{O}(\sigma,\phi^0)$.
Notice that now we are restricting the radial integration to the region $z \geq \epsilon > 0$.
This is necessary to avoid divergences in the integration.
Indeed, since $\Delta=d+1$ and $h_{ij}$ is quadratic in the scalar field, it follows that the backreaction goes as $1/z$ about the boundary, thus making the above integration divergent at $z=0$ and requiring that we integrate over the region $z \geq \epsilon > 0$.
This effect is reminiscent of~\cite{McGough:2016lol,Hartman:2018tkw}: the scalar field is causing the boundary to move into the bulk.
Finally, from form invariance of $G_{ij}$ we find to $\mathcal{O}(\sigma,\phi^2)$
\begin{align}\label{gmodrev}
    \delta g_{ij} + \delta h_{ij} &= 2\sigma (1-z\partial_z) \left( g_{ij} + h_{ij}\right)  \nonumber\\
    & \phantom{=} + \nabla_i a_j + \nabla_j a_i  + \nabla_i b_j + \nabla_j b_i + h_{ik} \nabla_j a^k + h_{jk} \nabla_i a^k + a^k \nabla_k h_{ij} \, ,
\end{align}
where indices are lowered with (and derivatives are covariant w.r.t.) $g_{ij}$.
We refer to eqs.~\eqref{modxi}, \eqref{gmodrev} as the modified PBH transformations.

Given the leading asymptotic behaviour of the metric and the backreaction, we make the following Ansätze for the radial expansion of $a^i$ and $b^i$
\begin{align}
    a^i(z,x) &= \overline{a}^i(\epsilon,x) + \sum_{n=1}^\infty z^n \, a^i_{(n)}(x) \\
    b^i (z,x) &= \overline{b}^i(\epsilon,x) +  \log z \, \Tilde{b}^i_{(1)}(x) + \sum_{n=2}^{\infty}\left[ z^{n-1} \left( \log z \, \Tilde{b}^i_{(n)}(x) + b^i_{(n)}(x) \right) \right] \, , \label{bansatz-d+1}
\end{align}
where $\overline{a}^i$ and $\overline{b}^i$ are constant terms in $z$ and their appearance is due to the lower end of the integration in~\eqref{modxi}.
For the metric and backreaction we assume
\begin{align}
    g_{ij}(z,x) &= \sum _{n=0}^\infty z^n \, g^{(n)}_{ij}(x) \\
    h_{ij}(z,x) &= \frac{1}{z} h^{(0)}_{ij}(x) + \log z \, \Tilde{h}^{(1)}_{ij}(x) + 0 + \sum_{n=2}^\infty \left[ z^{n-1} \left( \log z \, \Tilde{h}^{(n)}_{ij}(x) + h^{(n)}_{ij}(x) \right) \right] \, , \label{hansatz-d+1}
\end{align}
where we stress that there is no term at order $z^0$.
This implements the boundary condition that we anticipated above, namely that the boundary metric is still given by~$g^{(0)}_{ij}$ even in the presence of the backreaction~\cite{vanRees:hr_irr_op}.
The appearance of logarithmic terms is a consequence of the particular choice for the dimension of the scalar field $\Delta=d+1$, and we are thus generalising the analysis of~\cite{vanRees:hr_irr_op} as advocated in~\cite{vanRees:CS_eq-w_anomalies}.

Using the above expansions, from the modified PBH equations we find for the first few terms (the $a^i_{(n)}$ are as before)
\begin{align}
    \Tilde{b}^i_{(1)} =& - \frac{1}{2} g_{(0)}^{im}g_{(0)}^{jn} h^{(0)}_{mn} \partial_j \sigma \label{b1tilde} \\
    \Tilde{b}^i_{(2)} =& -\frac12 g_{(0)}^{im}g_{(0)}^{jn} \Tilde{h}^{(1)}_{mn} \partial_j \sigma \label{b2tilde}\\
    b^i_{(2)} =& \;  \frac{1}{2} \big[ \big( g_{(0)}^{im}g_{(1)}^{jn}  + g_{(1)}^{im}g_{(0)}^{jn} \big) h^{(0)}_{mn} + g_{(0)}^{im}g_{(0)}^{jn} \Tilde{h}^{(1)}_{mn}  \big] \partial_j \sigma \label{b2 d+1} \\
    \Tilde{b}^i_{(3)} =& \; \frac14 \big[  \big( g_{(0)}^{im}g_{(1)}^{jn}  + g_{(1)}^{im}g_{(0)}^{jn} \big) \Tilde{h}^{(1)}_{mn} - g_{(0)}^{im}g_{(0)}^{jn} \Tilde{h}^{(2)}_{mn}  \big] \partial_j \sigma \label{b3tilde}\\
    b^i_{(3)} =& -\frac{1}{4} \bigg[ h_{(2)}^{ij} -\frac12 \Tilde{h}_{(2)}^{ij} + \frac12 \big(  g_{(1)}^{ik} \Tilde{h}^{(1)j}{}_{k} + g_{(1)}^{jk} \Tilde{h}^{(1)i}{}_k \big) + g_{(1)}^{ik} g_{(1)}^{jl} h^{(0)}_{kl} + g_{(1)}^{ik} g^{(1)l}{}_k h^{(0)j}{}_l \nonumber \\
    & + g_{(1)}^{jk} g^{(1)l}{}_k  h^{(0)i}{}_l - g_{(2)}^{ik} h^{(0)j}{}_k - g_{(2)}^{jk} h^{(0)i}{}_k  \bigg] \partial_j \sigma \, , \label{b3 d+1}
\end{align}
where indices are lowered (raised) with the (inverse of) $g^{(0)}_{ij}$; curvatures and covariant derivatives will be w.r.t.~$g^{(0)}_{ij}$.
For the metric and backreaction variation\footnote{The expressions that follow are written up to boundary diffeomorphisms generated by $\bar{a}^i$ and $\bar{b}^i$. Since their presence does not affect the solution of the PBH equations, we disregard them for simplicity of notation.}
\begingroup
\allowdisplaybreaks[1]
\begin{align}
    \delta h^{(0)}_{ij} =& \; 4\sigma h^{(0)}_{ij} \label{deltah0 d+1} \\
    \delta \Tilde{h}^{(1)}_{ij} =& \; 2\sigma \Tilde{h}^{(1)}_{ij} + \overset{(0)}{\nabla}_i \Tilde{b}_{(1)j} + \overset{(0)}{\nabla}_j \Tilde{b}_{(1)i}  \label{deltah1tilde}\\
    \delta g^{(0)}_{ij} =& \; 2 \sigma g^{(0)}_{ij} - 2 \sigma \Tilde{h}^{(1)}_{ij} + h^{(0)}_{ik} \overset{(0)}{\nabla}_j a^k_{(1)} + h^{(0)}_{jk} \overset{(0)}{\nabla}_i a^k_{(1)} + a^k_{(1)} \overset{(0)}{\nabla}_k h^{(0)}_{ij} \label{gbeta}\\
    \vdots & \nonumber\\
    \delta \Tilde{h}^{(n+1)}_{ij} =& \; 2\sigma (1-n) \Tilde{h}^{(n+1)}_{ij} \nonumber\\*
    & + \sum_{m=1}^{n+1} \Big( g^{(n-m+1)}_{ik} \partial_j \tilde{b}^k_{(m)} + g^{(n-m+1)}_{jk} \partial_i \tilde{b}^k_{(m)} + \tilde{b}^k_{(m)} \partial_k g^{(n-m+1)}_{ij} \nonumber\\*
    & + \tilde{h}^{(n-m+1)}_{ik} \partial_j a^k_{(m)} + \tilde{h}^{(n-m+1)}_{jk} \partial_i a^k_{(m)} + a^k_{(m)} \partial_k \tilde{h}^{(n-m+1)}_{ij} \Big)\label{deltahntilde} \\
    \delta g^{(n)}_{ij} + \delta h^{(n+1)}_{ij} =& \, 2\sigma (1-n) \left( g^{(n)}_{ij} + h^{(n+1)}_{ij} \right) - 2\sigma \Tilde{h}^{(n+1)}_{ij} \nonumber\\*
    & + \sum_{m=1}^{n+1} \Big( g^{(n-m+1)}_{ik} \partial_j {b}^k_{(m)} + g^{(n-m+1)}_{jk} \partial_i {b}^k_{(m)} + {b}^k_{(m)} \partial_k g^{(n-m+1)}_{ij} \nonumber\\*
    & + {h}^{(n-m+1)}_{ik} \partial_j a^k_{(m)} + {h}^{(n-m+1)}_{jk} \partial_i a^k_{(m)} + a^k_{(m)} \partial_k {h}^{(n-m+1)}_{ij} \Big) \, . \label{deltag+hn}
\end{align}
\endgroup
A few comments are in order here.
The above equation for the metric and the backreaction can be solved in the same spirit outlined in sec.~\ref{sec:PBH}.
The term $h_{(n)}$ (and similarly $\tilde{h}_{(n)}$) is quadratic in $\phi_{(0)}$ and contains $2n$ derivatives.
Once the most general expression for $g_{(n)} + h_{(n+1)}$ (or $\tilde{h}_{(n)}$) is written down, it is enough to vary it according to~\eqref{gbeta} and~\eqref{dphi0} up to $\mathcal{O}(\sigma,\phi^2)$ and impose the variation is a PBH to find the sought for expression.
The solution of the backreacted Einstein's equations of motion (as in~\cite{vanRees:hr_irr_op} but with $\Delta = d+1$) will also satisfy the above equations.

As in the pure gravity case, the modified PBH equation fixes the expression of the backreaction only to some extent.
For instance, the first term in the expansion is
\begin{equation}\label{h0 d+1}
     h^{(0)}_{ij} = h_0 \, g^{(0)}_{ij} \phi_{(0)}^2
\end{equation}
for some coefficient $h_0$, not fixed by the PBH equation.
The higher order terms in the radial expansion will have more and more free coefficients, that are fixed on-shell given an action.\footnote{\label{foot:on-shell}For instance, given the action of a free massive scalar field coupled to a dynamical metric, then on-shell $h_0$ is proportional to the coefficient of the lowest order term in the radial expansion of the scalar field action and it is thus non-vanishing on-shell.}

Notice that the backreaction modifies the usual Weyl transformation of the boundary metric in~\eqref{gbeta}.
However, unlike in the pure gravity case, as it stands the modified PBH transformation does not reduce to a Weyl transformation of $g^{(0)}_{ij}$.
Indeed, when the diffeomorphism is a modified PBH transformation, from~\eqref{xihat} and~\eqref{modxi} we find that (up to $\mathcal O(\sigma,\phi^2)$)
\begin{equation}
    \hat\xi^z = 0 \, , \quad \hat\xi^i = \epsilon \big( g^{ij}(\epsilon,x) - h^{ij}(\epsilon,x) \big) \big( \xi^z_2 \partial_j \xi^z_1 - \xi^z_1 \partial_j \xi^z_2 \big) \, ,
\end{equation}
so that now we are left with a residual diffeomorphism
\begin{equation}\label{modPBH_comm}
    [\delta_2,\delta_1] g^{(0)}_{ij} = g^{(0)}_{jk}\overset{(0)}{\nabla}_i \hat\xi^k + g^{(0)}_{ik}\overset{(0)}{\nabla}_j \hat\xi^k \, .
\end{equation}
Before solving the modified PBH equation for the backreaction, we thus have to address the issue of the $z=\epsilon$ cutoff.

The holographic dual of the gravitational theory discussed so far is a CFT deformed by an irrelevant operator.
However, since we want to make contact with the unperturbed CFT presented in~\cite{Schwimmer:2019efk}, we have to move the cutoff surface back to the AdS boundary.
Given that 
\begin{equation}
    \bar{b}^i(\epsilon,x) = \frac12 h_0 \log \epsilon \, g_{(0)}^{ij} \phi_{(0)}^2 \partial_j \sigma + \mathcal{O}(\epsilon) \, ,
\end{equation}
the way to move the cutoff back to the boundary, without setting the source to zero, is to take $h_0 = 0$ and then $\epsilon = 0$.
In this limit we can describe a boundary CFT in the presence of an irrelevant operator, avoiding the prescription $\phi_{(0)}^2=0$ advocated in~\cite{Witten:1998qj,deHaro:2000vlm}.
As a bonus, the commutator in~\eqref{modPBH_comm} vanishes and the modified PBH transformation reduces to a Weyl transformation at the boundary.
However, the price to pay is that the solutions of the modified PBH equations are not on-shell anymore (see footnote~\ref{foot:on-shell}).

Now, we proceed by solving the modified PBH with $h_0 = 0$.
For $\tilde{h}_{(1)}$ we make the Ansatz
\begin{align}\label{h1tildeansatz}
    \tilde{h}^{(1)}_{ij} =& \;  h_1 \, \overset{(0)}{R}_{ij} \phi_{(0)}^2  + h_2 \, g^{(0)}_{ij} \overset{(0)}{R} \phi_{(0)}^2 + h_3 \, \overset{(0)}{\nabla}_i \phi_{(0)} \overset{(0)}{\nabla}_j \phi_{(0)} + h_4 \, \phi_{(0)} \overset{(0)}{\nabla}_i \overset{(0)}{\nabla}_j \phi_{(0)} \nonumber\\
    &\phantom{=}+ h_5 \, g^{(0)}_{ij} \phi_{(0)} \overset{(0)}{\Box} \phi_{(0)} + h_6 \, g^{(0)}_{ij} (\overset{(0)}{\nabla} \phi_{(0)})^2
\end{align}
and a solution of~\eqref{deltah1tilde} is given by
\begin{align}
    \Tilde{h}_{(1)ij} =& \; \, h_1 \left( \overset{(0)}{R}_{ij} \phi_{(0)}^2 + (d-2) \phi_{(0)} \overset{(0)}{\nabla}_i \overset{(0)}{\nabla}_j \phi_{(0)} + g^{(0)}_{ij} \phi_{(0)} \overset{(0)}{\Box} \phi_{(0)} - (d-1) g^{(0)}_{ij} ( \overset{(0)}{\nabla} \phi_{(0)} )^2  \right) \nonumber\\
    & + \, h_2 \, g^{(0)}_{ij} \left( \overset{(0)}{R} \phi_{(0)}^2 + 2(d-1) \phi_{(0)} \overset{(0)}{\Box} \phi_{(0)} -d(d-1) ( \overset{(0)}{\nabla} \phi_{(0)} )^2 \right) \, . \label{h1tilde}
\end{align}
Notice that the solutions parametrised by $h_1$ and $h_2$ are proportional to $\hat R_{ij}$ and $\hat R$ respectively, which are the curvature tensors computed from the Weyl-invariant metric $g^{(0)}_{ij}/\phi^2_{(0)}$.
Then, the variation of the metric in~\eqref{gbeta} reads
\begin{equation}\label{gbetanoh0}
    \delta g^{(0)}_{ij} = 2\sigma g^{(0)}_{ij} -2 \sigma \beta_{ij} \, , \quad \beta_{ij} \equiv h_1 \, \hat R_{ij} + h_2 \, g^{(0)}_{ij} \hat R
\end{equation}
and we interpret it as a modification of the usual Weyl transformation of the boundary metric due to the presence of the scalar field.
The holographic beta-function that we find
is in agreement with~\cite{Schwimmer:2019efk}.
Notice that for $h_1=0$ we could perform a field dependent redefinition of $\sigma$, so that the metric transformation is still a usual Weyl transformation:
\begin{equation}\label{sigma_redef}
    \sigma \to \hat\sigma = \sigma \left(1-h_2 \, \hat R \right) \quad \text{s.t.} \quad \delta g^{(0)}_{ij} = 2\hat\sigma g^{(0)}_{ij}
\end{equation}
We will comment on the effect of this redefinition to the holographic anomalies later.

Similarly, we also solve~\eqref{deltahntilde} for~$\tilde{h}_{(2)}$ and~\eqref{deltag+hn} for~$g_{(1)} + h_{(2)}$.
The Ansatz for~$h_{(2)}$ has thirty-five terms and the modified PBH equation leaves six out of the thirty-five coefficients free, while $\tilde{h}_{(2)}$ is determined in terms of $h_1$ and $h_2$.
We also solve for the trace of $\tilde{h}_{(3)}$ and the trace of $h_{(3)}$ that we will need in the following.
The Ansatz for the trace of~$h_{(3)}$ has sixty-six terms and the modified PBH equation leaves nine coefficients free, while the trace of $\tilde{h}_{(3)}$ is determined in terms of $h_1,h_2,c_1,c_2$.
$g_{(1)}$ and $g_{(2)}$ are not modified by the presence of the backreaction and are still given by~\eqref{g1} and~\eqref{g2} respectively.
We provide more details in app.~\ref{app:results} and in the ancillary Mathematica file.


\subsection{Effective boundary action and modified Weyl anomalies}
To derive the holographic dual of the modified Weyl anomaly found in~\cite{Schwimmer:2019efk}, we extend the method outlined in sec.~\ref{sec:PBHanomaly} to include the effect of the backreaction.

Consider an action
\begin{equation}\label{gravity+matter}
    S = \int_M d^{d+1} X \sqrt{G} f\left(R(G),\Phi\right) \, ,
\end{equation}
where $f$ is a local function of the curvature and its covariant derivatives and contains also matter field $\Phi$.
We could think of $S$ as gravitational action with a scalar field coupled to the metric and we require that $f\left(R(G), \Phi=0 \right)$ is such that $\text{AdS}_{d+1}$ is a solution of the equations of motion.
Then, we write the action in FG coordinates with backreaction~\eqref{modfg} and, expanding in powers of $z$, we obtain the holographic anomalies by evaluating the corresponding expressions on the solutions of the modified PBH equations.
However, the off-shell solution that we discussed so far with $h_0 = 0$ sets the scalar field action to zero (see footnote~\ref{foot:on-shell}), so that eventually the scalar field contributions to the anomalies is only due to the backreaction.

Defining now $\mathcal{L}=\sqrt{G} f\left(R(G),\Phi\right)$, in the FG coordinates~\eqref{modfg} we write
\begin{gather}
    \mathcal{L} = \sqrt{g_{(0)}} z^{-d/2-1} \left( \mathcal{L}_{g} + \mathcal{L}_{h} \right) \label{Og+Oh}\\
    \mathcal{L}_{g} = \sum_{n=0}^\infty \mathcal{L}^{(n)}_g (x) z^n \, , \quad \mathcal{L}_{h} = \sum_{n=0}^\infty \left[ \log z \, \tilde{\mathcal{L}}_h^{(n)} + \mathcal{L}_h^{(n)} \right] \, ,
\end{gather}
where $\mathcal{L}_h$ contains the backreaction and it is thus quadratic in the scalar field; in the second line we use the expansions~\eqref{hansatz-d+1} with $h_0 =0$ and $\mathcal{L}_g$ is as in sec.~\ref{sec:PBHanomaly}.
Following the reasoning of sec.~\ref{sec:PBHanomaly}, we can show that $[\delta_2,\delta_1]\mathcal{L}=0$ upon using the PBH group property for $h_0=0$ and thus $\mathcal{L}_{g}+\mathcal{L}_{h}$ satisfies the following Wess-Zumino condition
\begin{equation}\label{WZ_wbeta}
   \delta_1 \int_{\partial M} d^d x \sqrt{g_{(0)}} \, \sigma_2(x)  (\mathcal{L}_{g}+\mathcal{L}_{h}) - \delta_2 \int_{\partial M} d^d x \sqrt{g_{(0)}} \, \sigma_1(x) (\mathcal{L}_{g}+\mathcal{L}_{h}) = 0 \, ,
\end{equation}
where $g^{(0)}_{ij}$ now transforms with the beta-function~\eqref{gbetanoh0}.
From~\eqref{deltaSbdy} we interpret $\mathcal{L}_g^{(n)} + \mathcal{L}_h^{(n)}$ as the trace anomaly of the $d=2n$ dimensional CFT at the boundary.\footnote{As in sec.~\ref{sec:PBHanomaly}, we neglect divergent terms that are cancelled by counterterms.
In~\cite{deHaro:2000vlm} the counterterms considered only cancel negative powers of the radial coordinate; with irrelevant operators there is also need of logarithmic counterterms, as considered (in flat space) in~\cite{vanRees:CS_eq-w_anomalies}.}

Considering again the action~\eqref{EHwriem}, we now write it in the FG coordinates with backreaction~\eqref{modfg} and find the expression for $\mathcal{L}_h$ that we report in~\eqref{fwbackre} in app.~\ref{app:action}; $\mathcal{L}_g$ is still given by~\eqref{Og}.
Expanding the metric and backreaction in~\eqref{Og+Oh} according to~\eqref{hansatz-d+1} with $h_0 =0 $, we identify for instance
\begingroup
\allowdisplaybreaks[1]
\begin{align}
    \mathcal{L}_h^{(1)} &= 4\gamma \overset{(0)}{R}{}^{ij} \tilde{h}^{(1)}_{ij} + \frac12(2  - d) (1+4\gamma) h^{(2)i}{}_{i} - (2 - d) (1+8\gamma) g^{(1)ij} \tilde{h}^{(1)}{}_{ij} \nonumber\\*
    &\phantom{=} + \frac{1}{2} (1 +12\gamma - d(1+4\gamma)) g^{(1)i}{}_{i} \tilde{h}^{(1)j}{}_{j} + (3 - d) (1+4\gamma) \tilde{h}^{(2)i}{}_{i} \label{Oh1}\\
    \mathcal{L}_h^{(2)} &= - \frac{1}{4} (1 + \gamma ) \overset{(0)}{R} h^{(2)i}{}_{i} + 2 (1 + 6 \gamma - \frac{3}{8} d (1 + 4 \gamma )) g^{(1)j}{}_{j} h^{(2)i}{}_{i} + \frac{1}{2}(1 + 12 \gamma ) \overset{(0)}{R}{}_{ij} h^{(2)ij} \nonumber\\
    &\phantom{=} - (5 +28 \gamma  - \frac12 d (3 + 20 \gamma )) g^{(1)}{}_{ij} h^{(2)ij} - \frac{3}{2} (d-4) (1 + 4 \gamma ) h^{(3)i}{}_{i}  \nonumber\\ 
    &\phantom{=} - (5  +68 \gamma  -d(1+ 12\gamma) ) g^{(2)ij} \tilde{h}^{(1)}{}_{ij} -8 \gamma  \overset{(0)}{R}{}^{ij} g^{(1)}{}_{i}{}^{k} \tilde{h}^{(1)}{}_{jk} + 8 \gamma  \nabla_{k}\tilde{h}^{(1)}{}_{ij} \nabla^{k}g^{(1)ij} \nonumber\\
    &\phantom{=} + (5 + 56 \gamma   - d (1 + 12 \gamma )) g^{(1)}{}_{i}{}^{k} g^{(1)ij} \tilde{h}^{(1)}{}_{jk} -8 \gamma  \overset{(0)}{\nabla}_j\tilde{h}^{(1)}{}_{ik}\overset{(0)}{\nabla}{}^{k}g^{(1)ij}  \nonumber\\
    &\phantom{=} - \frac{1}{2} (4  +48 \gamma   -d(1+ 8 \gamma )) g^{(1)i}{}_{i} g^{(1)jk} \tilde{h}^{(1)}{}_{jk} + \frac{1}{2} (3 + 28 \gamma  - d (1 + 4 \gamma )) g^{(2)i}{}_{i} \tilde{h}^{(1)j}{}_{j} \nonumber\\
    &\phantom{=} - \frac{1}{4} (3 +44 \gamma  -d(1+ 4 \gamma) ) g^{(1)}{}_{ij} g^{(1)ij} \tilde{h}^{(1)k}{}_{k} +\frac{1}{2} ( 1+ 4 \gamma ) \overset{(0)}{\Box}h^{(2)i}{}_{i}  \nonumber\\
    &\phantom{=} + \frac{1}{8} (3 + 28 \gamma  - d (1 + 4 \gamma )) g^{(1)i}{}_{i} g^{(1)j}{}_{j} \tilde{h}^{(1)k}{}_{k} -4 \gamma  \overset{(0)}{R}_{ikjl} g^{(1)ij} \tilde{h}^{(1)kl} + 4 \gamma  \overset{(0)}{R}{}^{ij} \tilde{h}^{(2)}{}_{ij} \nonumber\\
    &\phantom{=} - (6  +32 \gamma  +d(1+ 8 \gamma) ) g^{(1)ij} \tilde{h}^{(2)}{}_{ij} + \frac{1}{2} (5 + 28 \gamma  - d (1 + 4 \gamma )) g^{(1)i}{}_{i} \tilde{h}^{(2)j}{}_{j} \nonumber\\
    &\phantom{=}  -2 \gamma  \tilde{h}^{(1)ij} \overset{(0)}{\nabla}_{j}\overset{(0)}{\nabla}_{i}g^{(1)k}{}_{k}  + 4 \gamma  \tilde{h}^{(1)ij} \nabla_{k}\nabla_{j}g^{(1)}{}_{i}{}^{k} - \frac{1}{2} (1 +4 \gamma ) \overset{(0)}{\nabla}_{j}\overset{(0)}{\nabla}_{i}h^{(2)ij} \nonumber\\
    &\phantom{=}- (d-7) (1 + 4 \gamma ) \tilde{h}^{(3)i}{}_{i} + 2 \gamma  \overset{(0)}{R}{}^{ij} g^{(1)k}{}_{k} \tilde{h}^{(1)}{}_{ij} -2 \gamma  \tilde{h}^{(1)ij} \overset{(0)}{\Box}g^{(1)}{}_{ij}\label{Oh2}
\end{align}
\endgroup
which should contribute to the anomaly in $d=2,4$ respectively.
Focusing on the $d=4$ case, the candidate anomaly is $\mathcal{L}_g^{(2)} + \mathcal{L}_h^{(2)}$, which on the PBH solutions yields indeed a solution of the Wess-Zumino condition.
Notice that $g^{(n)}$ appears in $\mathcal{L}_h^{(n)}$ and together with~$\tilde{h}^{(3)i}{}_{i}$ it causes factors of $(d-4)^{-1}$ to appear in $\mathcal{L}_h^{(2)}$.
Thus, at first sight $\mathcal{L}_g^{(2)} + \mathcal{L}_h^{(2)}$ is singular in $d=4$.
Nonetheless, it is possible to renormalise the free coefficients of the PBH solution for the backreaction (see app.~\ref{app:coeffren}) so that eventually $\mathcal{L}_g^{(2)} + \mathcal{L}_h^{(2)}$ is regular in $d=4$ and can be identified with the holographic anomaly of the boundary CFT.
We thus define $\mathcal{A}^{4d} = (\mathcal{L}_g^{(2)} + \mathcal{L}_h^{(2)})|_{\text{reg}}$ as the regularised, four-dimensional holographic anomaly, which of course still satisfies the Wess-Zumino condition.
The term quadratic in the curvature in~\eqref{EHwriem} allows us to separate again the pure gravity type A and B anomalies, which now receive contributions also from the scalar field.
In $\mathcal{A}^{4d}$ we then identify
\begin{equation}\label{full_ano}
    \mathcal{A}^{4d} = \mathcal{A}^{4d}_{\text{B}} + \mathcal{A}^{4d}_{\text{A}}
\end{equation}
with
\begin{align}
    \mathcal{A}^{4d}_{\text{B}} &= \mathcal{I}^{\text{pg}}_\text{B} + \mathcal{I}_1(h_1) + \mathcal{I}_2 (c_1,h_1,h_2) + \mathcal{I}_3(c_2,h_1,h_2) \label{full_ano_B}\\
    \mathcal{A}^{4d}_{\text{A}} &= \mathcal{I}^{\text{pg}}_\text{A} + \mathcal{I}_4(h_1) + \mathcal{I}_5(h_2) + \mathcal{I}_6(h_2) + \mathcal{I}_7(c_1,h_1,h_2) + \mathcal{I}_8(c_2,h_1,h_2) \nonumber\\
    &\phantom{=} + \mathcal{I}_9(h_{21}) + \mathcal{I}_{10}(h_{22}) + \mathcal{I}_{11}(h_{23})+ \mathcal{I}_{12}(h_{24}) + \mathcal{I}_{13}(h_{25}) + \mathcal{I}_{14}(h_{26}) \label{full_ano_A}
\end{align}
where $\mathcal{I}^{\text{pg}}_\text{B} = - c\, C^2 /16 $ and $\mathcal{I}^{\text{pg}}_\text{A} = a\, E_4 /16$ are the type B and A pure gravity anomaly and the $\mathcal{I}_i$ are tensorial structures quadratic in the scalar field and to sixth order in derivatives, parametrised by the (backreaction) coefficients given in parenthesis.
In particular, $\mathcal{I}_i$ for $i=1,2,3$ contribute to the type B pure gravity anomaly, while $\mathcal{I}_i$ for $i=4,\ldots,14$ contribute to the type A pure gravity anomaly.
We present all the explicit expressions in appendix~\ref{app:full_ano} (and for convenience also in the ancillary Mathematica notebook);
here, we discuss the general properties of the individual tensorial structures and make contact with the results of~\cite{Schwimmer:2019efk}.

As we already remarked, $\mathcal{A}^{4d}$ satisfies the Wess-Zumino condition~\eqref{WZ_wbeta} to $\mathcal{O}(\phi^2)$ when the metric transforms with the beta-function~\eqref{gbetanoh0}.
However, some of the tensorial structures in~$\mathcal{A}^{4d}$ are solutions of the Wess-Zumino condition without need of the beta-function.
Amongst these solutions, we can distinguish between trivial solutions (i.e.~terms which in the field theory can be cancelled with counterterms) and solutions which corresponds to `ordinary' anomaly in the CFT that cannot be cancelled with counterterms.
The $\mathcal{I}_i$ for $i=9,\ldots,14$ are trivial solutions, while $\mathcal{I}_i$ for $i=2,3,6,7,8$ correspond to ordinary anomalies.
Notice that, even though the latter depend on the metric beta-function coefficients, they are solutions of the Wess-Zumino condition without need of the beta-function, so that from the field theory point of view they do not correspond to modifications of the type A and B anomalies induced by the beta-function.
The structures $\mathcal{I}_i$ for $i=2,3,7,8$ transform under Weyl transformations as $\delta \mathcal{I}_i = -4\sigma \mathcal{I}_i$, so that they lead to Weyl-invariant anomaly integrals; nonetheless, $\mathcal{I}_7$ and $\mathcal{I}_8$ fall into the scalar field contributions to the type A pure gravity anomaly and thus we deduce that the bulk action~\eqref{EHwriem} does not have enough parameters to distinguish between the type A and B terms quadratic in the scalar field.
To do so, it might be necessary to include additional higher-derivative terms in the bulk action, which would complicate the computation of the anomalies and for simplicity we didn't consider.

The remaining structures $\mathcal{I}_i$ for $i=1,4,5$ together with the pure gravity terms are solutions of the Wess-Zumino condition only when the beta-function is taken into account.
However, as we mentioned above, for $h_1 =0$ we can perform a redefinition of $\sigma$ so that the transformation~\eqref{gbetanoh0} is a usual Weyl transformation~\eqref{sigma_redef}.
This suggests that from the field theory point of view the anomaly seeded by the beta-function parametrised by $h_2$ might not be a new anomaly, but rather a deformation of the usual pure gravity $\mathcal{I}^{\text{pg}}_\text{A}$ anomaly.
Indeed, from our holographic computation we find that
\begin{equation}
    \mathcal{I}_5(h_2) = -\frac{a}{16} h_2 \hat R \, E_4
\end{equation}
so that 
\begin{equation}
    \sigma \left( \mathcal{I}^{\text{pg}}_\text{A} + \mathcal{I}_5(h_2) \right) = \frac{a}{16} \hat \sigma E_4
\end{equation}
which is a solution of the Wess-Zumino condition when the metric transforms in the usual way with a redefined $\sigma$ as in~\eqref{sigma_redef}.

The tensorial structures $\mathcal{I}_1$ and $\mathcal{I}_4$, together with the pure gravity terms, are the only terms which survive in~\eqref{full_ano} when all the free coefficients appearing in the anomaly, but $h_1$, are set to zero.
We now focus on these terms to make contact with the anomalies in~\cite{Schwimmer:2019efk}.
Setting to zero all the free coefficients in the anomaly, but $h_1$, we are then left with the following expression for the pure gravity type B anomaly and its scalar field contributions ($g_{ij} \equiv g^{(0)}_{ij}$ and $\phi \equiv \phi_{(0)}$):
\begingroup
\allowdisplaybreaks[1]
\begin{align}
    \mathcal{A}^{4d}_{\text{B}} &= \mathcal{I}^{\text{pg}}_\text{B} + \mathcal{I}_1(h_1) \nonumber\\
    &= -\frac{1}{16} c \, C^2 + c\, h_1 \Bigg( \frac{3}{4} \nabla^{j}\nabla^{i}\phi \Box\nabla_{j}\nabla_{i}\phi -  \frac{3}{16} \Box\phi \Box^2 \phi + \frac{1}{2} \nabla_{k}\nabla_{j}\nabla_{i}\phi \nabla^{k}\nabla^{j}\nabla^{i}\phi \nonumber\\
    &\phantom{=} - \frac{89}{144} R_{i}{}^{k} R^{ij} R_{jk} \phi^2 + \frac{97}{144} R_{ij} R^{ij} R \phi^2 -  \frac{7}{81} R^3 \phi^2 -  \frac{11}{16} R^{ij} R^{kl} R_{ikjl} \phi^2 \nonumber\\
    &\phantom{=} -  \frac{5}{384} R R_{ijkl} R^{ijkl} \phi^2 -  \frac{1}{18} R_{i}{}^{m}{}_{k}{}^{n} R^{ijkl} R_{jmln} \phi^2 + \frac{53}{576} R_{ij}{}^{mn} R^{ijkl} R_{klmn} \phi^2 \nonumber\\
    &\phantom{=} -  \frac{7}{192} R \phi^2 \Box R + \frac{3}{64} R^2 \phi \Box \phi + \frac{1}{6} R_{jklm} R^{jklm} \phi \Box \phi -  \frac{19}{288} \phi^2 \nabla_{i}R \nabla^{i}R \nonumber\\
    &\phantom{=} -  \frac{65}{288} R \phi \nabla_{i}\phi \nabla^{i}R + R^{jk} \phi \nabla_{i}R_{jk} \nabla^{i}\phi + \frac{1}{16} R^{jklm} \phi \nabla_{i}R_{jklm} \nabla^{i}\phi \nonumber\\
    &\phantom{=} + \frac{19}{24} R_{jk} R^{jk} \nabla_{i}\phi \nabla^{i}\phi -  \frac{5}{36} R^2 \nabla_{i}\phi \nabla^{i}\phi -  \frac{1}{3} R_{jklm} R^{jklm} \nabla_{i}\phi \nabla^{i}\phi \nonumber\\
    &\phantom{=} -  \frac{1}{36} R^{ij} \phi^2 \nabla_{j}\nabla_{i}R -  \frac{11}{72} R^{ij} R \phi \nabla_{j}\nabla_{i}\phi -  \frac{35}{288} \phi \Box R \Box \phi -  \frac{13}{48} \nabla_{i}\phi \nabla^{i}R \Box \phi \nonumber\\
    &\phantom{=} -  \frac{1}{12} \phi \nabla^{i}R \Box \nabla_{i}\phi -  \frac{1}{6} R \nabla^{i}\phi \Box \nabla_{i}\phi -  \frac{3}{32} R \phi \Box^2\phi + \frac{1}{3} R_{ij} \phi \nabla^{i}R \nabla^{j}\phi \nonumber\\
    &\phantom{=} + \frac{1}{12} R_{ij} R \nabla^{i}\phi \nabla^{j}\phi -  \frac{1}{2} R^{kl} R_{ikjl} \nabla^{i}\phi \nabla^{j}\phi + \frac{1}{12} \nabla^{i}R \nabla_{j}\nabla_{i}\phi \nabla^{j}\phi \nonumber\\
    &\phantom{=} -  \frac{1}{18} \phi \nabla_{j}\nabla_{i}\phi \nabla^{j}\nabla^{i}R -  \frac{3}{16} R \nabla_{j}\nabla_{i}\phi \nabla^{j}\nabla^{i}\phi -  \frac{1}{2} R^{jk} \phi \nabla^{i}\phi \nabla_{k}R_{ij} \nonumber\\
    &\phantom{=} -  \frac{1}{8} R_{i}{}^{k} R^{ij} \phi \nabla_{k}\nabla_{j}\phi + \frac{1}{2} R^{jk} \nabla^{i}\phi \nabla_{k}\nabla_{j}\nabla_{i}\phi + \frac{1}{8} R^{ij} \phi^2 \Box R_{ij} + \frac{13}{24} \phi \nabla^{j}\nabla^{i}\phi \Box R_{ij} \nonumber\\
    &\phantom{=} -  \frac{1}{6} R_{ij} R^{ij} \phi \Box \phi + \frac{1}{2} R_{i}{}^{j} \nabla^{i}\phi \Box \nabla_{j}\phi + \frac{3}{8} R^{ij} \phi \Box \nabla_{j}\nabla_{i}\phi -  \frac{9}{16} \phi^2 \nabla_{j}R_{ik} \nabla^{k}R^{ij} \nonumber\\
    &\phantom{=} + \frac{61}{96} \phi^2 \nabla_{k}R_{ij} \nabla^{k}R^{ij} + \frac{1}{2} \phi \nabla_{k}\nabla_{j}\nabla_{i}\phi \nabla^{k}R^{ij} + \frac{3}{8} R^{ij} \nabla_{k}\nabla_{j}\phi \nabla^{k}\nabla_{i}\phi \nonumber\\
    &\phantom{=} + \frac{5}{4} \nabla_{i}R_{jk} \nabla^{i}\phi \nabla^{k}\nabla^{j}\phi -  \frac{1}{2} \nabla^{i}\phi \nabla_{k}R_{ij} \nabla^{k}\nabla^{j}\phi + \frac{1}{2} R_{ijkl} \phi \nabla^{i}\phi \nabla^{l}R^{jk} \nonumber\\
    &\phantom{=} + \frac{7}{24} R_{ikjl} \phi^2 \nabla^{l}\nabla^{k}R^{ij} + \frac{5}{24} R^{ij} R_{ikjl} \phi \nabla^{l}\nabla^{k}\phi + \frac{3}{8} R_{ikjl} \nabla^{j}\nabla^{i}\phi \nabla^{l}\nabla^{k}\phi \nonumber\\
    &\phantom{=} + \frac{5}{192} \phi^2 \nabla_{m}R_{ijkl} \nabla^{m}R^{ijkl} \Bigg) \, , \label{modetypeB_h1}
\end{align}
\endgroup
while the pure gravity type A anomaly and the scalar field contributions read (again $g_{ij} \equiv g^{(0)}_{ij}$ and $\phi \equiv \phi_{(0)}$):
\begingroup
\allowdisplaybreaks[1]
\begin{align}
    \mathcal{A}^{4d}_{\text{A}} &= \mathcal{I}^{\text{pg}}_\text{A} + \mathcal{I}_4(h_1) \nonumber\\
    &= \frac{1}{16} a\, E_4 + a\, h_1 \Bigg(    \frac{2}{3} \nabla^{j}\Box\phi \Box\nabla_{j}\phi -  \frac{43}{24} \nabla^{j}\nabla^{i}\phi \Box \nabla_{j}\nabla_{i}\phi + \frac{31}{96} \Box\phi \Box^2\phi \nonumber\\
    &\phantom{=} -  \frac{1}{4} \nabla^{i}\phi \Box^2\nabla_{i}\phi + \frac{7}{32} \phi \Box^3\phi -  \frac{5}{3} \nabla_{k}\nabla_{j}\nabla_{i}\phi \nabla^{k}\nabla^{j}\nabla^{i}\phi + \frac{43}{96} R_{i}{}^{k} R^{ij} R_{jk} \phi^2 \nonumber\\
    &\phantom{=} + \frac{1}{12} R_{ij} R^{ij} R \phi^2 -  \frac{23}{768} R^3 \phi^2 -  \frac{53}{96} R^{ij} R^{kl} R_{ikjl} \phi^2 -  \frac{29}{768} R R_{ijkl} R^{ijkl} \phi^2 \nonumber\\
    &\phantom{=} + \frac{41}{24} R_{i}{}^{m}{}_{k}{}^{n} R^{ijkl} R_{jmln} \phi^2 -  \frac{61}{384} R_{ij}{}^{mn} R^{ijkl} R_{klmn} \phi^2 + \frac{1}{288} R \phi^2 \Box R -  \frac{85}{144} R^2 \phi \Box \phi \nonumber\\
    &\phantom{=} -  \frac{19}{48} R_{jklm} R^{jklm} \phi \Box \phi -  \frac{7}{576} \phi^2 \nabla_{i}R \nabla^{i}R -  \frac{19}{36} R \phi \nabla_{i}\phi \nabla^{i}R + 4 R^{jk} \phi \nabla_{i}R_{jk} \nabla^{i}\phi \nonumber\\
    &\phantom{=} -  \frac{169}{48} R^{jklm} \phi \nabla_{i}R_{jklm} \nabla^{i}\phi + \frac{155}{32} R_{jk} R^{jk} \nabla_{i}\phi \nabla^{i}\phi -  \frac{295}{576} R^2 \nabla_{i}\phi \nabla^{i}\phi \nonumber\\
    &\phantom{=} -  \frac{535}{192} R_{jklm} R^{jklm} \nabla_{i}\phi \nabla^{i}\phi + \frac{5}{8} R^{ij} \phi^2 \nabla_{j}\nabla_{i}R + \frac{77}{48} R^{ij} R \phi \nabla_{j}\nabla_{i}\phi \nonumber\\
    &\phantom{=} -  \frac{1}{16} \nabla_{i}\phi \nabla^{i}\phi \Box R + \frac{55}{96} \phi \Box R \Box\phi + \frac{17}{48} R \Box\phi \Box\phi + \frac{23}{96} \nabla_{i}\phi \nabla^{i}R \Box\phi \nonumber\\
    &\phantom{=} + \frac{1}{4} \phi \nabla^{i}\phi \Box \nabla_{i}R + \frac{55}{96} \phi \nabla^{i}R \Box\nabla_{i}\phi -  \frac{5}{6} R \nabla^{i}\phi \Box \nabla_{i}\phi + \frac{7}{192} \phi^2 \Box^2 R \nonumber\\
    &\phantom{=} + \frac{13}{32} R \phi \Box^2 \phi -  \frac{49}{24} R_{ij} \phi \nabla^{i}R \nabla^{j}\phi  + \frac{13}{6} R_{i}{}^{k} R_{jk} \nabla^{i}\phi \nabla^{j}\phi - 2 R_{ij} R \nabla^{i}\phi \nabla^{j}\phi \nonumber\\
    &\phantom{=} + \frac{13}{6} R^{kl} R_{ikjl} \nabla^{i}\phi \nabla^{j}\phi -  \frac{7}{24} \nabla^{i}\phi \nabla_{j}\nabla_{i}R \nabla^{j}\phi + \frac{67}{48} \phi \nabla_{j}\nabla_{i}\phi \nabla^{j}\nabla^{i}R \nonumber\\
    &\phantom{=} -  \frac{51}{32} R \nabla_{j}\nabla_{i}\phi \nabla^{j}\nabla^{i}\phi + \frac{69}{16} R^{jk} \phi \nabla^{i}\phi \nabla_{k}R_{ij} + \frac{17}{16} R_{i}{}^{k} R^{ij} \phi \nabla_{k}\nabla_{j}\phi \nonumber\\
    &\phantom{=} + \frac{37}{12} R^{jk} \nabla^{i}\phi \nabla_{k}\nabla_{j}\nabla_{i}\phi + \frac{5}{16} R^{ij} \phi^2 \Box R_{ij} + \frac{3}{8} \nabla^{i}\phi \nabla^{j}\phi \Box R_{ij} -  \frac{139}{48} \phi \nabla^{j}\nabla^{i}\phi \Box R_{ij} \nonumber\\
    &\phantom{=} + \frac{21}{8} R_{ij} R^{ij} \phi \Box \phi -  \frac{5}{24} R^{ij} \nabla_{j}\nabla_{i}\phi \Box \phi -  \frac{29}{12} R_{i}{}^{j} \nabla^{i}\phi \Box \nabla_{j}\phi -  \frac{13}{48} R^{ij} \phi \Box \nabla_{j}\nabla_{i}\phi \nonumber\\
    &\phantom{=} + \frac{185}{64} \phi^2 \nabla_{j}R_{ik} \nabla^{k}R^{ij} -  \frac{153}{128} \phi^2 \nabla_{k}R_{ij} \nabla^{k}R^{ij} + \frac{1}{12} \phi \nabla_{k}\nabla_{j}\nabla_{i}\phi \nabla^{k}R^{ij} \nonumber\\
    &\phantom{=} + \frac{27}{16} R^{ij} \nabla_{k}\nabla_{j}\phi \nabla^{k}\nabla_{i}\phi -  \frac{31}{24} \nabla_{i}R_{jk} \nabla^{i}\phi \nabla^{k}\nabla^{j}\phi -  \nabla^{i}\phi \nabla_{k}R_{ij} \nabla^{k}\nabla^{j}\phi \nonumber\\
    &\phantom{=} -  \frac{41}{8} R_{ijkl} \phi \nabla^{i}\phi \nabla^{l}R^{jk} -  \frac{223}{48} R_{ikjl} \phi^2 \nabla^{l}\nabla^{k}R^{ij} -  \frac{101}{16} R^{ij} R_{ikjl} \phi \nabla^{l}\nabla^{k}\phi \nonumber\\
    &\phantom{=} -  \frac{13}{16} R_{ikjl} \nabla^{j}\nabla^{i}\phi \nabla^{l}\nabla^{k}\phi -  \frac{643}{768} \phi^2 \nabla_{m}R_{ijkl} \nabla^{m}R^{ijkl}  \Bigg) \, . \label{modtypeA_h1}
\end{align}
\endgroup
The expressions~\eqref{modetypeB_h1} and~\eqref{modtypeA_h1} satisfy the Wess-Zumino condition~\eqref{WZ_wbeta} and, since they are parametrised by $a,c$ and $h_1$ only, we can compare them to the anomalies presented in~\cite{Schwimmer:2019efk}.
At first sight, they look different from the one obtained in~\cite{Schwimmer:2019efk}.
However, we checked that the expressions are the same,\footnote{Up to a factor of 2, which is missing in the normalisation of the metric beta-function in~\cite{Schwimmer:2019efk}.\label{foot:normalisation}} up to variation of local counterterms in the field theory and addition of ordinary Weyl-invariant anomalies.
These terms are present in~\eqref{modetypeB_h1} and~\eqref{modtypeA_h1} and the bulk action~\eqref{EHwriem} does not have enough parameters to distinguish them  from the modifications of the pure gravity anomalies which are solutions of~\eqref{WZ_wbeta} only when the beta-function is taken into account.

Notice the interplay between the $a$, $c$ coefficients and the metric beta-function coefficient $h_1$, as already observed in~\cite{Schwimmer:2019efk}.
It is this feature that makes the anomalies~\eqref{modetypeB_h1} and~\eqref{modtypeA_h1} solutions of the consistency condition~\eqref{WZ_wbeta} in the presence of the metric beta-function.
We remark that the other tensorial structures which appear in the anomaly solve the Wess-Zumino condition without need of the metric beta-function (apart from $\mathcal{I}_5$, whose peculiarities we analysed above).

As a final comment, we discuss the flat space limit of~\eqref{full_ano}.
From the explicit expressions in app.~\ref{app:full_ano}, it appears that in flat boundary space and for constant $\sigma\to 1$, the anomaly reduces to (after integration by parts)
\begin{equation}\label{flat}
    \int d^4 x \sqrt{g_{(0)}} \, \sigma(x) \mathcal{A}^{4d} \to \frac{1}{16} c \, h_1 \, \int d^4 x \, \phi_{(0)} \Box^3 \phi_{(0)} \, ,
\end{equation}
i.e.~the type A anomaly vanishes in flat space and the result is proportional to $h_1$.
This contribution comes entirely from~$\mathcal{I}_1$.
The conformal anomaly associated with free massive scalar fields in flat boundary space and its coefficient are computed holographically in~\cite{deHaro:2000vlm}.
Comparing with our result, we can fix the coefficient $h_1$ as
\begin{equation}
    h_1 = \frac{16}{c} \frac{1}{128} \, .
\end{equation}
Notice that this coefficient does not renormalise~\cite{Petkou:1999fv}.
The flat space limit~\eqref{flat} and the value of $h_1$ are in agreement with the results in~\cite{Schwimmer:2019efk}.
Indeed, comparing with~\eqref{gbetanoh0}, from~\cite{Schwimmer:2019efk} we identify (with footnote~\ref{foot:normalisation} in mind)
\begin{equation}
    h_1 = \frac{5}{72} \frac{N}{c_T}
\end{equation}
where $N$ is the normalisation of the two-point function of scalar operators which according to~\cite{Freedman:1998tz} is $N= \pi^{-d/2}(2\Delta-d)\Gamma(\Delta)/\Gamma(\Delta-d/2)$ and from the normalisation of the type B Weyl anomaly we find $c_T = 40 c / \pi^2$.
For $d=4$ and $\Delta =5$, the value of $h_1$ computed in the CFT matches the one found holographically.

\section{Conclusions}
In this paper we computed the holographic Weyl anomaly in the presence of a source for irrelevant operators of the boundary theory, when the boundary theory is a four-dimensional CFT.
In the bulk, the massive scalar field induces a backreaction onto the metric, and, since the scalar field sources irrelevant operator on the boundary, it changes the leading asymptotic behaviour of the metric which becomes more singular as the boundary is approached.
This causes logarithmic terms to appear in the radial expansion of the backreaction which affect the Weyl transformation of the boundary metric.
In particular, a metric beta-function appears, which in turn modifies the holographic Weyl anomaly.
The metric beta-function and the holographic Weyl anomaly are in agreement with the known CFT results, so that the present analysis provides an additional test of the AdS/CFT correspondence.

To obtain the holographic results, we studied diffeomorphisms in the bulk and used the equivalence between bulk diffeomorphisms and Weyl transformation at the boundary.
We saw that the presence of a scalar field, which sources an irrelevant operator at the boundary, introduces a radial cutoff, which reflects the fact that the boundary field theory is being deformed by the irrelevant operator.
Eliminating the cutoff, we were able to describe the underformed CFT and the modified PBH transformations reduce to a boundary Weyl transformation.
However, this requires that the backreaction is put off-shell.
It would be interesting to extend the present analysis by considering an on-shell backreaction, so that the dual theory is a CFT deformed by irrelevant operators.
The metric beta-function would then indicate that the geometry is subject to an RG flow, and it would be interesting to provide its holographic description.

The holographic anomaly is obtained as the finite piece of a bulk action, evaluated on the PBH solutions.
The scalar field bulk action does not contribute to the holographic anomaly, which is obtained by the gravitational action alone evaluated on the PBH solution for the metric including the backreaction due to the scalar field.
In the resulting expression, in addition to the usual pure gravity Weyl anomaly there are contributions which depend on the scalar field and we calculated them to lowest non-trivial order, which is quadratic in $\phi_{(0)}$.
Some of them correspond to anomalous terms seeded by the metric beta-function, while others correspond to solutions of the Wess-Zumino consistency condition to quadratic order in the scalar field but not seeded by the backreaction.
The holographic anomaly is non-vanishing in flat space, where it reduces to the expected result, which in $d$ boundary dimensions for a scalar operator of dimension $\Delta$ with source $\phi_{(0)}$ is proportional to $\phi_{(0)} \Box^n \phi_{(0)}$ with integer $n = \Delta - d/2$~\cite{vanRees:CS_eq-w_anomalies}.

The present analysis could be generalised to higher-dimensional theories, where the calculations will be more complicated but similar to the one we presented here.
However, it does not provide the correct dual description of a two-dimensional boundary CFT.
Indeed, for $d=2$ and $\Delta=3$ the candidate holographic anomaly should be given by $\mathcal{L}_g^{(1)} + \mathcal{L}_h^{(1)}$ with $\mathcal{L}_h^{(1)}$ given by~\eqref{Oh1}.
On the PBH solutions it yields an expression which satisfies the Wess-Zumino condition~\eqref{WZ_wbeta}, but it vanishes in flat space and therefore does not reproduce the expected result which should be proportional to $\phi_{(0)} \Box^2 \phi_{(0)}$.
This happens because for $d=2$ the solutions parametrised by $h_1$ and $h_2$ in~\eqref{h1tilde} are no longer independent.
The breakdown of the present description in $d=2$ is expected, as the CFT analysis in two dimensions~\cite{Schwimmer:2019efk} requires the addition of a $\Delta=4$ operator, which is eventually identified with the $T \bar T$ operator, together with an operator of dimension $\Delta =3$.
We checked that including an additional scalar field in the bulk which sources a $\Delta =4$ operator on the boundary and then demanding that the boundary CFT is not deformed by the irrelevant operators still does not reproduce the $d=2$ metric beta-function found in~\cite{Schwimmer:2019efk}.
This may be because the standard holographic dictionary tells that the $T \bar T$ operator is dual to a sharp cutoff in the bulk.
However, we saw that requiring that the dual CFT is undeformed translated into eliminating the cutoff induced by the irrelevant operator. 
It would thus be interesting to extend the present analysis to the two-dimensional case and see how the CFT results are reproduced by holographic computations.

\section*{Acknowledgements}
I gratefully acknowledge the original suggestion of this project by Stefan Theisen, and helpful discussions and critical comments to the draft of this paper.
I also thank Hugo Camargo, Lorenzo Casarin, Ruben Manvelyan and Adam Schwimmer for discussions and comments.
I am supported by the International Max Planck Research School for Mathematical and Physical Aspects of Gravitation, Cosmology and Quantum Field Theory.


\appendix

\section{Ansätze}\label{app:results}

In this appendix we collect the Ans\"atze that we make to solve eqs.~\eqref{deltahntilde},~\eqref{deltag+hn} for $n=1,2$.
We report the explicit results in the ancillary Mathematica notebook.\footnote{The results were obtained with the help of the xAct collection of Mathematica packages~\cite{Martin-Garcia:2007bqa,Martin-Garcia:2008yei,MARTINGARCIA2008597,Brizuela:2008ra,Nutma:2013zea}.}
In the following, we assume that $g_{ij} \equiv g^{(0)}_{ij}$ and curvatures and covariant derivatives are w.r.t.~$g^{(0)}_{ij}$, and $\phi \equiv \phi_{(0)}$.
We also choose $h_0 =0$, as explained in the text.

The solution of the PBH equation for $g^{(1)}_{ij}$ is given in~\eqref{g1} and it is not modified by the presence of the backreaction.
The most generic Ansatz for $h^{(2)}_{ij}$ has thirty-five terms and six coefficients will not be fixed by the modified PBH equation.
We write the Ansatz as
\begingroup
\allowdisplaybreaks[1]
\begin{align}
   h^{(2)}_{ij} =& \;  + h_{21}^{} g_{ij} R_{kl} R^{kl} \phi^2 + h_{22}^{} R_{ij} R \phi^2 + h_{23}^{} R^{kl} R_{ikjl} \phi^2 + h_{24}^{} \phi^2 \nabla_{j}\nabla_{i}R  + h_{25}^{} R_{ij} \phi \Box \phi \nonumber\\
   &\phantom{=} + h_{26}^{} g_{ij} R \phi \Box \phi  + \beta_{1}^{} R_{i}{}^{k} R_{jk} \phi^2  + \beta_{2}^{} g_{ij} R^2 \phi^2  + \beta_{3}^{} R_{i}{}^{klm} R_{jklm} \phi^2 \nonumber\\
   &\phantom{=} + \beta_{4}^{} g_{ij} R_{klmn} R^{klmn} \phi^2 + \beta_{6}^{} R \nabla_{i}\phi \nabla_{j}\phi + \beta_{5}^{} (\phi \nabla_{i}\phi \nabla_{j}R + \phi \nabla_{i}R \nabla_{j}\phi) \nonumber\\
   &\phantom{=} + \beta_{7}^{} R \phi \nabla_{j}\nabla_{i}\phi + \beta_{8}^{} (R_{j}{}^{k} \phi \nabla_{k}\nabla_{i}\phi + R_{i}{}^{k} \phi \nabla_{k}\nabla_{j}\phi) + \beta_{9}^{} \phi^2 \Box R_{ij} + \beta_{10}^{} g_{ij} \phi^2 \Box R \nonumber\\
   &\phantom{=} + \beta_{11}^{} \nabla_{j}\nabla_{i}\phi \Box \phi + \beta_{12}^{} (\nabla_{j}\phi \Box \nabla_{i}\phi + \nabla_{i}\phi \Box \nabla_{j}\phi) + \beta_{13}^{} \phi \Box\nabla_{j}\nabla_{i}\phi \nonumber\\
   &\phantom{=} + \beta_{14}^{} g_{ij} \phi \nabla_{k}\phi \nabla^{k}R + \beta_{15}^{} (\phi \nabla_{i}R_{jk} \nabla^{k}\phi + \phi \nabla_{j}R_{ik} \nabla^{k}\phi) \nonumber\\
   &\phantom{=} + \beta_{16}^{} (R_{jk} \nabla_{i}\phi \nabla^{k}\phi + R_{ik} \nabla_{j}\phi \nabla^{k}\phi) + \beta_{17}^{} \phi \nabla_{k}R_{ij} \nabla^{k}\phi + \beta_{18}^{} R_{ij} \nabla_{k}\phi \nabla^{k}\phi \nonumber\\
   &\phantom{=} + \beta_{19}^{} g_{ij} R \nabla_{k}\phi \nabla^{k}\phi + \beta_{20}^{} \nabla_{k}\nabla_{j}\nabla_{i}\phi \nabla^{k}\phi + \beta_{21}^{} \nabla_{k}\nabla_{j}\phi \nabla^{k}\nabla_{i}\phi \nonumber\\
   &\phantom{=} + \beta_{22}^{} g_{ij} R^{kl} \phi \nabla_{l}\nabla_{k}\phi  + \beta_{23}^{} g_{ij} \Box \phi \Box \phi + \beta_{24}^{} g_{ij} \nabla^{k}\phi \Box \nabla_{k}\phi + \beta_{25}^{} g_{ij} \phi \Box^2 \phi \nonumber\\
   &\phantom{=} + \beta_{26}^{} g_{ij} R_{kl} \nabla^{k}\phi \nabla^{l}\phi + \beta_{27}^{} R_{ikjl} \nabla^{k}\phi \nabla^{l}\phi + \beta_{28}^{} R_{ikjl} \phi \nabla^{l}\nabla^{k}\phi \nonumber\\
   &\phantom{=} + \beta_{29}^{} g_{ij} \nabla_{l}\nabla_{k}\phi \nabla^{l}\nabla^{k}\phi 
\end{align}
\endgroup
and leave the coefficients $h_{21},\ldots,h_{26}$ free.
The full solution is in the ancillary notebook.

A similar Ansatz is made for $\tilde{h}^{(2)}_{ij}$ and its coefficients are fixed in terms of $h_1$ and $h_2$ by the modified PBH equation.

The solution for $g^{(2)}_{ij}$ is given in~\eqref{g2} and it is not modified by the presence of the backreaction.
Thus, here we focus on ${h}^{(3)}_{ij}$, which appears in eq.~\eqref{Oh2} only as ${h}^{(3)i}{}_{i}$ and we then do not compute ${h}^{(3)}_{ij}$ but rather its trace that satisfies the following PBH equation 
\begin{equation}
    \delta g^{(2)i}{}_{i} + \delta {h}^{(3)i}{}_{i} = -2\sigma g^{(2)i}{}_{i} + g^{(0)ij} \delta g^{(2)}{}_{ij} +2\sigma \beta^{ij} g^{(2)}_{ij} -2\sigma {h}^{(3)i}{}_{i} + g^{(0)ij} \delta {h}^{(3)}_{ij} \, .
\end{equation}
The most generic Ansatz for ${h}^{(3)i}{}_{i}$ has sixty-six terms and nine coefficients will not be fixed by the modified PBH equation.
We write the Ansatz as
\begingroup
\allowdisplaybreaks[1]
\begin{align}
    {h}^{(3)i}{}_{i} =& \; h_{31}^{} R^{jk} \phi \nabla_{i}R_{jk} \nabla^{i}\phi + h_{32}^{} R^{ij} \phi^2 \nabla_{j}\nabla_{i}R + h_{33}^{} R_{ij} R^{ij} R \phi^2 + h_{34}^{} R^{ij} R \phi \nabla_{j}\nabla_{i}\phi \nonumber\\
    &\phantom{=} + h_{35}^{} R_{ij} \phi \nabla^{i}R \nabla^{j}\phi + h_{36}^{} R^{ij} R^{kl} R_{ikjl} \phi^2 + h_{37}^{} R^{ij} \phi^2 \Box R_{ij} + h_{38}^{} R_{ij} R^{ij} \phi \Box \phi \nonumber\\
    &\phantom{=} + h_{39}^{} R^{ij} \nabla_{j}\nabla_{i}\phi \Box \phi + \gamma_{1}^{} R_{i}{}^{k} R^{ij} R_{jk} \phi^2  + \gamma_{2}^{} R^3 \phi^2  + \gamma_{3}^{} R R_{ijkl} R^{ijkl} \phi^2 \nonumber\\
    &\phantom{=} + \gamma_{4}^{} R^{ij} R_{i}{}^{klm} R_{jklm} \phi^2 + \gamma_{5}^{} R_{i}{}^{m}{}_{k}{}^{n} R^{ijkl} R_{jmln} \phi^2 + \gamma_{6}^{} R_{ij}{}^{mn} R^{ijkl} R_{klmn} \phi^2 \nonumber\\
    &\phantom{=} + \gamma_{7}^{} R \phi^2 \Box R + \gamma_{8}^{} R^2 \phi \Box \phi + \gamma_{9}^{} R_{jklm} R^{jklm} \phi \Box \phi + \gamma_{10}^{} \phi^2 \nabla_{i}R \nabla^{i}R \nonumber\\
    &\phantom{=} + \gamma_{11}^{} R \phi \nabla_{i}\phi \nabla^{i}R  + \gamma_{12}^{} R^{jklm} \phi \nabla_{i}R_{jklm} \nabla^{i}\phi + \gamma_{13}^{} R_{jk} R^{jk} \nabla_{i}\phi \nabla^{i}\phi \nonumber\\
    &\phantom{=} + \gamma_{14}^{} R^2 \nabla_{i}\phi \nabla^{i}\phi + \gamma_{15}^{} R_{jklm} R^{jklm} \nabla_{i}\phi \nabla^{i}\phi   + \gamma_{16}^{} \nabla_{i}\phi \nabla^{i}\phi \Box R + \gamma_{17}^{} \phi \Box R \Box \phi \nonumber\\
    &\phantom{=} + \gamma_{18}^{} R \Box \phi \Box \phi + \gamma_{19}^{} \nabla_{i}\phi \nabla^{i}R \Box\phi + \gamma_{20}^{} \phi \nabla^{i}\phi \Box \nabla_{i}R + \gamma_{21}^{} \phi \nabla^{i}R \Box \nabla_{i}\phi \nonumber\\
    &\phantom{=} + \gamma_{22}^{} R \nabla^{i}\phi \Box \nabla_{i}\phi + \gamma_{23}^{} \phi^2 \Box^2 R + \gamma_{24}^{} R \phi \Box^2\phi  + \gamma_{25}^{} R_{i}{}^{k} R_{jk} \nabla^{i}\phi \nabla^{j}\phi \nonumber\\
    &\phantom{=} + \gamma_{26}^{} R_{ij} R \nabla^{i}\phi \nabla^{j}\phi + \gamma_{27}^{} R^{kl} R_{ikjl} \nabla^{i}\phi \nabla^{j}\phi + \gamma_{28}^{} R_{i}{}^{klm} R_{jklm} \nabla^{i}\phi \nabla^{j}\phi \nonumber\\
    &\phantom{=} + \gamma_{29}^{} \nabla^{i}\phi \nabla_{j}\nabla_{i}R \nabla^{j}\phi + \gamma_{30}^{} \nabla^{i}R \nabla_{j}\nabla_{i}\phi \nabla^{j}\phi + \gamma_{31}^{} \phi \nabla_{j}\nabla_{i}\phi \nabla^{j}\nabla^{i}R \nonumber\\
    &\phantom{=} + \gamma_{32}^{} R_{i}{}^{klm} R_{jklm} \phi \nabla^{j}\nabla^{i}\phi + \gamma_{33}^{} R \nabla_{j}\nabla_{i}\phi \nabla^{j}\nabla^{i}\phi + \gamma_{34}^{} R^{jk} \phi \nabla^{i}\phi \nabla_{k}R_{ij} \nonumber\\
    &\phantom{=} + \gamma_{35}^{} R_{i}{}^{k} R^{ij} \phi \nabla_{k}\nabla_{j}\phi + \gamma_{36}^{} R^{jk} \nabla^{i}\phi \nabla_{k}\nabla_{j}\nabla_{i}\phi  + \gamma_{37}^{} \nabla^{i}\phi \nabla^{j}\phi \Box R_{ij} \nonumber\\
    &\phantom{=} + \gamma_{38}^{} \phi \nabla^{j}\nabla^{i}\phi \Box R_{ij}  + \gamma_{39}^{} R_{i}{}^{j} \nabla^{i}\phi \Box \nabla_{j}\phi + \gamma_{40}^{} \nabla^{j}\Box \phi \Box \nabla_{j}\phi \nonumber\\
    &\phantom{=} + \gamma_{41}^{} R^{ij} \phi \Box \nabla_{j}\nabla_{i}\phi  + \gamma_{42}^{} \nabla^{j}\nabla^{i}\phi \Box\nabla_{j}\nabla_{i}\phi + \gamma_{43}^{} \Box\phi \Box^2\phi + \gamma_{44}^{} \nabla^{i}\phi \Box^2\nabla_{i}\phi \nonumber\\
    &\phantom{=} + \gamma_{45}^{} \phi \Box^3\phi  + \gamma_{46}^{} \phi^2 \nabla_{j}R_{ik} \nabla^{k}R^{ij} + \gamma_{47}^{} \phi^2 \nabla_{k}R_{ij} \nabla^{k}R^{ij} \nonumber\\
    &\phantom{=} + \gamma_{48}^{} \phi \nabla_{k}\nabla_{j}\nabla_{i}\phi \nabla^{k}R^{ij}  + \gamma_{49}^{} R^{ij} \nabla_{k}\nabla_{j}\phi \nabla^{k}\nabla_{i}\phi + \gamma_{50}^{} \nabla_{i}R_{jk} \nabla^{i}\phi \nabla^{k}\nabla^{j}\phi \nonumber\\
    &\phantom{=} + \gamma_{51}^{} \nabla^{i}\phi \nabla_{k}R_{ij} \nabla^{k}\nabla^{j}\phi  + \gamma_{52}^{} \nabla_{k}\nabla_{j}\nabla_{i}\phi \nabla^{k}\nabla^{j}\nabla^{i}\phi + \gamma_{53}^{} R_{ijkl} \phi \nabla^{i}\phi \nabla^{l}R^{jk} \nonumber\\
    &\phantom{=} + \gamma_{54}^{} R_{ikjl} \phi^2 \nabla^{l}\nabla^{k}R^{ij}  + \gamma_{55}^{} R^{ij} R_{ikjl} \phi \nabla^{l}\nabla^{k}\phi + \gamma_{56}^{} R_{ikjl} \nabla^{j}\nabla^{i}\phi \nabla^{l}\nabla^{k}\phi \nonumber\\
    &\phantom{=} + \gamma_{57}^{} \phi^2 \nabla_{m}R_{ijkl} \nabla^{m}R^{ijkl}\label{h3ansatz}
\end{align}
\endgroup
and leave the coefficients $h_{31},\ldots,h_{39}$ free.
The full solution is in the ancillary notebook.

Similarly, since $\tilde{h}^{(3)}$ appears in eq.~\eqref{Oh2} only as $\tilde{h}^{(3)i}{}_{i}$, we do not compute $\tilde{h}^{(3)}_{ij}$ but rather its trace that satisfies the following PBH equation 
\begin{equation}
    \delta \tilde{h}^{(3)i}{}_{i} = -2\sigma \tilde{h}^{(3)i}{}_{i} + g^{(0)ij} \delta \tilde{h}^{(3)}_{ij} \, .
\end{equation}
A similar Ansatz to~\eqref{h3ansatz} is made for $\tilde{h}^{(3)i}{}_{i}$ and its coefficients are fixed in terms of $h_1,h_2,c_1,c_2$ by the modified PBH equation.

\section{Gravitational action with backreaction}\label{app:action}
From the action~\eqref{EHwriem} in FG coordinates with backreaction~\eqref{modfg} we identify:
\begingroup
\allowdisplaybreaks[1]
\begin{align}
     \mathcal{L}_h &= \tfrac12 d(1 + 4 \gamma ) h^{i}{}_{i} \nonumber\\
     & + \tfrac12 z (1+4\gamma)\big[ 2 (d-1) g'_{ij} h^{ij} - (d-1) g^{ij}g'_{ij} h^{k}{}_{k} \nonumber\\
     &\phantom{=} -2 (d-1) g^{ij}h'_{ij} + h^{ij} R_{ij} - \tfrac{1}{2} h^{i}{}_{i} R -\nabla^{j}\nabla^{i}h_{ij}  + \Box h^{i}{}_{i}\big] \nonumber\\
     & + z^2 \big[ -2 (1 + 4\gamma ) g''_{ij} h^{ij} + (3  -4 (d-5) \gamma ) g^{im}g^{jn}g'_{ik} g'_{mn} h^{jk} \nonumber\\
     &\phantom{=}- (1 + 8 \gamma ) g^{ij} g'_{ij} g'_{kl} h^{kl} + (1 + 4 \gamma ) g^{ij} g''_{ij} h^{k}{}_{k} -2 \gamma  R_{ikjl} \nabla^{l}\nabla^{k}h^{ij} \nonumber\\
     &\phantom{=}- (\tfrac{3}{4} -  (d-5) \gamma ) g^{im}g^{jn}g'_{ij} g'_{mn} h^{k}{}_{k} + (\tfrac{1}{4} + 2 \gamma ) (g^{ij}g'_{ij})^2 h^{k}{}_{k} \nonumber\\
     &\phantom{=}- (3 - 4 (d-5) \gamma ) g^{ik}g^{jl}g'_{ij} h'_{kl} + 2(1 + 4 \gamma ) g^{ij} h''_{ij} + 2 \gamma  g'_{ij}  R^{ij} h^{k}{}_{k} \nonumber\\
     &\phantom{=} + (1 + 8 \gamma ) g^{ij} g^{kl} g'{ij} h'_{kl} + 4 \gamma h'_{ij} R^{ij} -8 \gamma  g'_{ij} h^{i}{}_{k} R^{jk} - \gamma  h^{ij} R_{i}{}^{klp} R_{jklp} \nonumber\\
     &\phantom{=} + \tfrac{1}{4} \gamma  h^{i}{}_{i} R_{jklp} R^{jklp} -2 \gamma  g'_{ij} \nabla^{j}\nabla^{i}h^{k}{}_{k} + 4 \gamma  g'_{ij} \nabla_{k}\nabla^{j}h^{ik} -2 \gamma  g'_ {ij} \Box h^{ij} \big] \nonumber\\
     & + \gamma z^3 \big[ -12 g^{im}g^{jn} g'_{ik} g'_{mn} g'_{jl} h^{kl} + 8 g^{ij} g^{mn} g'_{ij} g'_{lm} g'_{mk} h^{kl} \nonumber\\
     &\phantom{=}+ 4 g^{im}g^{jn} g'_{ij} g'_{mn} g'_{kl} h^{kl} + 2 g^{im}g^{jn}g^{kl} g'_{ik}g'_{mn} g'_{jl} h^{p}{}_{p} \nonumber\\
     &\phantom{=} -2 g^{ij} g^{km} g^{nl} g'_{ij} g'_{kl} g'_{mn} h^{p}{}_{p}  + 12 g^{im}g^{jn} g^{kl} g'_{ik} g'_{mn} h'_{jl} \nonumber\\
     &\phantom{=}-8 g^{ij} g^{km} g^{ln} g'_{ij} g'_{kl} h'_{mn} -4 g^{ik} g^{jl} g^{mn} g'_{ij} g'_{kl} h'_{mn}  - g'_{ij} g'_{kl} h^{p}{}_{p} R^{ikjl} \nonumber\\
     &\phantom{=} -4 g'_{ij} h'_{kl} R^{ikjl} -6 g'_{ij} g'_{kl} h^{i}{}_{p} R^{jklp} -4 g^{km} g^{ln} h^{ij} \nabla_{i}g'_{kl} \nabla_{j}g'_{mn} \nonumber\\
     &\phantom{=} -4 g'_{ij} \nabla^{i}g'_{kl} \nabla^{j}h^{kl} -8 g^{il} \nabla^{j}h'_{ik} \nabla^{k}g'_{jl} + 8 g^{ik}g^{jl}\nabla_{k}h'_{ij} \nabla^{k}g'_{kl} \nonumber\\
     &\phantom{=} + 8 g'_{ij} \nabla^{i}g'_{kl} \nabla^{l}h^{jk} -2  g'_{ij} g'_{kl} \nabla^{l}\nabla^{j}h^{ik} + 2 g'^{ij} g'^{kl} \nabla^{l}\nabla^{k}h^{ij}  \nonumber\\
     &\phantom{=} + 8 g^{kn} h^{ij} \nabla_{j}g'_{kl} \nabla^{l}g'_{in} + 4 g^{in} g'_{jn} \nabla^{j}h^{kl} \nabla_{l}g'_{ik} + 4 h^{ij} \nabla^{k}g'_{jl} \nabla^{l}g'_{ik} \nonumber\\
     &\phantom{=} -4 g^{in} g'_{ij} \nabla^{k}h^{jl} \nabla_{l}g'_{kn} -8 g^{kn} h^{ij} \nabla_{l}g'_{jk} \nabla^{l}g'_{in} -4 g^{in} g'_{ij} \nabla_{l}h^{jk} \nabla^{l}g'_{kn} \nonumber\\
     &\phantom{=} -2 g^{jn} h^{i}{}_{i} \nabla^{k}g'_{jl} \nabla^{l}g'_{kn} + 2 g^{jm}g^{km} h^{i}{}_{i} \nabla_{l}g'_{jk} \nabla^{l}g'_{mn}\big] \nonumber\\
     & +\gamma z^4 \big[ 8 g^{im}g^{kn}g'_{ij} g'_{kl} g''_{mn} h^{jl} -16 g^{il}g''_{ik} g''_{jl} h^{jk}  + 16 g^{im} g^{jn} g'_{ik} g'_{mn} g''_{jl} h^{kl} \nonumber\\
     &\phantom{=} + 4 g^{ik} g^{jl}  g''_{ij} g''_{kl} h^{k}{}_{k} -4  g^{kq} g^{ir} g^{js} g'_{ik} g'_{rs} g'_{jl} g'_{pq} h^{lp} -8 g^{im}g^{jn}g^{kl} g'_{ik}g'_{mn} h''_{jl} \nonumber\\
     &\phantom{=} -4  g^{ik}g^{jl} g^{mr}g^{ns}g^{pq} g'_{ij} g'_{kl} g'_{mn} g'_{rq} h^{ps} -4 g^{im}g^{jn}g^{kl} g'_{ik}g'_{mn} g''_{jl} h^{p}{}_{p} \nonumber\\
     &\phantom{=} + \tfrac12 g^{im}g^{jn} g^{kr} g^{ls} g'_{ik} g'_{mn} g'_{jl} g'_{rs} h^{p}{}_{p}  + \tfrac12 (g^{ik}g^{jl}g'_{ij} g'_{kl})^2 h^{p}{}_{p} -16 g^{im}g^{jn} g^{kl} g'_{mn} g''_{ik} h'_{jl}\nonumber\\
     &\phantom{=} + 4 g^{im}g^{jn}g^{kr}g^{ls} g'_{ik} g'_{mn} g'_{jl} h'_{rs} + 4 g^{im}g^{jn} g^{kr} g^{ls} g'_{ij} g'_{mn} g'_{kl} h'_{rs} + 16 g^{ik}g^{jl}g''_{ij} h''_{kl} \big] \label{fwbackre}
\end{align}
\endgroup
while $\mathcal{L}_g$ is still given by~\eqref{Og}.
Notice that also terms of $\mathcal{O}(z^4)$ contribute to the anomaly in $d=4$, since $h''_{ij} = - z^{-2}\tilde{h}^{(1)}_{ij} + \mathcal{O}(z^{-1})$.

\section{Coefficients renormalisation}\label{app:coeffren}
Evaluating~\eqref{Og2} and~\eqref{Oh2} on the PBH solutions, we find that $\mathcal{L}_g^{(2)} + \mathcal{L}_h^{(2)}$ satisfies the Wess-Zumino condition~\eqref{WZ_wbeta}, even though it is singular in $d=4$.
In particular, expanding around $d=4$ we find that
\begin{equation}\label{og2oh2series}
    \mathcal{L}_g^{(2)} + \mathcal{L}_h^{(2)} = \frac{1}{d-4} I^{(-1)} + I^{(0)} + (d-4) I^{(1)} + (d-4)^2 I^{(2)} + \ldots
\end{equation}
with ($g_{ij} \equiv g^{(0)}_{ij}$ and $\phi \equiv \phi_{(0)}$)
\begin{align}
    I^{(-1)} &= c \, h_{1} \Bigg( \frac{7}{24}  R_{ij} R^{ij} R \phi^2 -  \frac{1}{24} R^3 \phi^2 -  \frac{1}{2} R^{ij} R^{kl} R_{ikjl} \phi^2 + \frac{1}{24} R \phi^2 \Box R  -  \frac{1}{12} R^2 \phi \Box\phi \nonumber\\
    &\phantom{=}+ \frac{1}{12} R^{ij} \phi^2 \nabla_{j}\nabla_{i}R + \frac{1}{3} R^{ij} R \phi \nabla_{j}\nabla_{i}\phi + \frac{1}{12} \phi \Box R \Box\phi  + \frac{1}{6} \phi \nabla_{j}\nabla_{i}\phi \nabla^{j}\nabla^{i}R \nonumber\\
    &\phantom{=} -  \frac{1}{4}  R^{ij} \phi^2 \Box R_{ij} -  \frac{1}{2}  \phi \nabla^{j}\nabla^{i}\phi \Box R_{ij} + \frac{1}{4}  R_{ij} R^{ij} \phi \Box\phi -  R^{ij} R_{ikjl} \phi \nabla^{l}\nabla^{k}\phi \Bigg) \, .
\end{align}
Notice that $I^{(-1)}$ is entirely type B and $\delta I^{(-1)} = -4\sigma I^{(-1)}$ under Weyl transformation up to $\mathcal{O}(\phi^2)$.
However, $\mathcal{L}_g^{(2)} + \mathcal{L}_h^{(2)}$ contains also several free coefficients, which are not fixed by the PBH equations, i.e.~$\{c_1,c_2,h_1,h_{21},\ldots,h_{26},h_{31},\ldots,h_{39}\}$, and it is possible to regularise~\eqref{og2oh2series} by renormalising some of these coefficients.
In particular, there is a unique shift which cancels $I^{(-1)}$ and it consists of shifting some of the free coefficients which appear in $I^{(2)}$ as follows:
\begin{alignat*}{4}
    & h_{32} \to h_{32} + \frac{c \, h_1}{18 a (d-4)^2} \quad && h_{33} \to h_{33} + \frac{7 c \, h_1}{36 a (d-4)^2} \quad && h_{34} \to h_{34} + \frac{2 c\, h_1}{9a (d-4)^2} \\
    & h_{36} \to h_{36} - \frac{c\, h_1}{3 a (d-4)^2} \quad && h_{37} \to h_{37} - \frac{c\, h_1}{6 a (d-4)^2} \quad && h_{38} \to h_{38} + \frac{c\, h_1}{6a (d-4)^2} \; .
\end{alignat*}
Since the shifted coefficients only appear in $I^{(2)}$, this shift does not introduce new singularities, and of course the regularised expression satisfies the Wess-Zumino condition.

\section{Complete expression of the anomaly}\label{app:full_ano}
We provide here the complete expression of the holographic anomaly $\mathcal{A}^{4d}$ as in~\eqref{full_ano}, recalling that in the type A and B pure gravity anomalies with the respective scalar field contributions we identify the following terms
\begin{align}
    \mathcal{A}^{4d}_{\text{B}} &= \mathcal{I}^{\text{pg}}_\text{B} + \mathcal{I}_1(h_1) + \mathcal{I}_2 (c_1,h_1,h_2) + \mathcal{I}_3(c_2,h_1,h_2) \label{full_ano_B_app}\\
    \mathcal{A}^{4d}_{\text{A}} &= \mathcal{I}^{\text{pg}}_\text{A} + \mathcal{I}_4(h_1) + \mathcal{I}_5(h_2) + \mathcal{I}_6(h_2) + \mathcal{I}_7(c_1,h_1,h_2) + \mathcal{I}_8(c_2,h_1,h_2) \nonumber\\
    &\phantom{=} + \mathcal{I}_9(h_{21}) + \mathcal{I}_{10}(h_{22}) + \mathcal{I}_{11}(h_{23})+ \mathcal{I}_{12}(h_{24}) + \mathcal{I}_{13}(h_{25}) + \mathcal{I}_{14}(h_{26}) \, . \label{full_ano_A_app}
\end{align}
We proceed by presenting and analysing them individually.
Again, we assume that $g_{ij} \equiv g^{(0)}_{ij}$ and curvatures and covariant derivatives are w.r.t.~$g^{(0)}_{ij}$, and $\phi \equiv \phi_{(0)}$.
For convenience, we report the expressions of $\mathcal{A}^{4d}_{\text{B}}$ and $\mathcal{A}^{4d}_{\text{A}}$ also in the ancillary Mathematica notebook.

As for $\mathcal{A}^{4d}_{\text{B}}$, we already presented  $\mathcal{I}^{\text{pg}}_\text{B}$ and $\mathcal{I}_1$ in~\eqref{modetypeB_h1}.
We move then to $\mathcal{I}_2$, which is given by
\begin{equation}
    \mathcal{I}_2 (c_1,h_1,h_2) = 4 c \, c_1 (h_1+4 h_2) \hat R \, C^2 \, .
\end{equation}
This tensorial structure leads to a Weyl-invariant anomaly integrand, as already pointed out in~\cite{Schwimmer:2019efk}.
Thus, in the CFT it contributes to the anomaly as an `ordinary' anomaly, which satisfy the Wess-Zumino condition without need of the beta-function.
Henceforth we will refer to this kind of structures as `ordinary anomaly' for short.
Notice that $\mathcal{I}_2$ vanishes in flat space.
A similar discussion follows for $\mathcal{I}_3$ since it is related to $\mathcal{I}_2$ by
\begin{equation}
    \mathcal{I}_3 (c_2,h_1,h_2) = \mathcal{I}_2\left(\frac{c_2}{4},h_1,h_2\right) \, .
\end{equation}

As for $\mathcal{A}^{4d}_{\text{A}}$, the tensorial structures $\mathcal{I}^{\text{pg}}_\text{A}$ and $\mathcal{I}_4$ are given in~\eqref{modtypeA_h1} and $\mathcal{I}_5$ is also already presented and discussed in the main text.
Then, we move to $\mathcal{I}_6$ which corresponds to ordinary anomalies and thus satisfies the WZ condition without need of the metric beta-function.
Explicitly, it reads
\begingroup
\allowdisplaybreaks[1]
\begin{align}
    \mathcal{I}_6(h_2) &= a \, h_2 \Big( 3 \nabla^{j}\Box\phi \Box\nabla_{j}\phi -  \tfrac{15}{2} \nabla^{j}\nabla^{i}\phi \Box\nabla_{j}\nabla_{i}\phi + \tfrac{15}{8} \Box\phi \Box^2\phi + \tfrac{9}{8} \phi \Box^2\phi \nonumber\\
    &\phantom{=}-  \tfrac{15}{2} \nabla_{k}\nabla_{j}\nabla_{i}\phi \nabla^{k}\nabla^{j}\nabla^{i}\phi - \tfrac{3}{8} R_{i}{}^{k} R^{ij} R_{jk} \phi^2 + \tfrac{5}{2} R_{ij} R^{ij} R \phi^2 -  \tfrac{23}{64} R^3 \phi^2 \nonumber\\
    &\phantom{=} -  \tfrac{39}{8} R^{ij} R^{kl} R_{ikjl} \phi^2 -  \tfrac{17}{64} R R_{ijkl} R^{ijkl} \phi^2 + \tfrac{33}{4} R_{i}{}^{m}{}_{k}{}^{n} R^{ijkl} R_{jmln} \phi^2 \nonumber\\
    &\phantom{=} -  \tfrac{9}{32} R_{ij}{}^{mn} R^{ijkl} R_{klmn} \phi^2 -  \tfrac{3}{16} R \phi^2 \Box R -  \tfrac{17}{8} R^2 \phi \Box \phi -  \tfrac{15}{8} R_{jklm} R^{jklm} \phi \Box \phi \nonumber\\
    &\phantom{=} -  \tfrac{1}{2} \phi^2 \nabla_{i}R \nabla^{i}R -  \tfrac{15}{4} R \phi \nabla_{i}\phi \nabla^{i}R + \tfrac{99}{4} R^{jk} \phi \nabla_{i}R_{jk} \nabla^{i}\phi \nonumber\\
    &\phantom{=} -  \tfrac{129}{8} R^{jklm} \phi \nabla_{i}R_{jklm} \nabla^{i}\phi + \tfrac{189}{8} R_{jk} R^{jk} \nabla_{i}\phi \nabla^{i}\phi -  \tfrac{23}{8} R^2 \nabla_{i}\phi \nabla^{i}\phi \nonumber\\
    &\phantom{=} -  \tfrac{51}{4} R_{jklm} R^{jklm} \nabla_{i}\phi \nabla^{i}\phi + \tfrac{19}{8} R^{ij} \phi^2 \nabla_{j}\nabla_{i}R + 6 R^{ij} R \phi \nabla_{j}\nabla_{i}\phi + \tfrac{17}{8} \phi \Box R \Box \phi \nonumber\\
    &\phantom{=} + \tfrac{17}{8} R \Box \phi \Box \phi + \tfrac{1}{8} \nabla_{i}\phi \nabla^{i}R \Box \phi + \tfrac{3}{2} \phi \nabla^{i}\phi \Box\nabla_{i}R + \tfrac{21}{8} \phi \nabla^{i}R \Box\nabla_{i}\phi \nonumber\\
    &\phantom{=} -  \tfrac{15}{4} R \nabla^{i}\phi \Box\nabla_{i}\phi + \tfrac{3}{16} \phi^2 \Box^2 R + \tfrac{3}{2} R \phi \Box^2 \phi -  \tfrac{27}{4} R_{ij} \phi \nabla^{i}R \nabla^{j}\phi \nonumber\\
    &\phantom{=} + \tfrac{27}{2} R_{i}{}^{k} R_{jk} \nabla^{i}\phi \nabla^{j}\phi -  \tfrac{37}{4} R_{ij} R \nabla^{i}\phi \nabla^{j}\phi + \tfrac{15}{2} R^{kl} R_{ikjl} \nabla^{i}\phi \nabla^{j}\phi \nonumber\\
    &\phantom{=} + \tfrac{7}{4} \nabla^{i}R \nabla_{j}\nabla_{i}\phi \nabla^{j}\phi + \tfrac{23}{4} \phi \nabla_{j}\nabla_{i}\phi \nabla^{j}\nabla^{i}R -  \tfrac{59}{8} R \nabla_{j}\nabla_{i}\phi \nabla^{j}\nabla^{i}\phi \nonumber\\
    &\phantom{=} + \tfrac{57}{4} R^{jk} \phi \nabla^{i}\phi \nabla_{k}R_{ij} + \tfrac{15}{4} R_{i}{}^{k} R^{ij} \phi \nabla_{k}\nabla_{j}\phi + \tfrac{27}{2} R^{jk} \nabla^{i}\phi \nabla_{k}\nabla_{j}\nabla_{i}\phi \nonumber\\
    &\phantom{=} + \tfrac{9}{4} R^{ij} \phi^2 \Box R_{ij} -  \tfrac{39}{4} \phi \nabla^{j}\nabla^{i}\phi \Box R_{ij} + \tfrac{21}{2} R_{ij} R^{ij} \phi \Box\phi -  \tfrac{15}{4} R^{ij} \nabla_{j}\nabla_{i}\phi \Box\phi \nonumber\\
    &\phantom{=} - 9 R_{i}{}^{j} \nabla^{i}\phi \Box\nabla_{j}\phi + \tfrac{129}{16} \phi^2 \nabla_{j}R_{ik} \nabla^{k}R^{ij} -  \tfrac{9}{32} \phi^2 \nabla_{k}R_{ij} \nabla^{k}R^{ij} \nonumber\\
    &\phantom{=} + \tfrac{3}{2} \phi \nabla_{k}\nabla_{j}\nabla_{i}\phi \nabla^{k}R^{ij} + \tfrac{45}{4} R^{ij} \nabla_{k}\nabla_{j}\phi \nabla^{k}\nabla_{i}\phi + 3 \nabla_{i}R_{jk} \nabla^{i}\phi \nabla^{k}\nabla^{j}\phi \nonumber\\
    &\phantom{=} -  \tfrac{21}{2} \nabla^{i}\phi \nabla_{k}R_{ij} \nabla^{k}\nabla^{j}\phi - 18 R_{ijkl} \phi \nabla^{i}\phi \nabla^{l}R^{jk} -  \tfrac{75}{4} R_{ikjl} \phi^2 \nabla^{l}\nabla^{k}R^{ij} \nonumber\\
    &\phantom{=} -  \tfrac{99}{4} R^{ij} R_{ikjl} \phi \nabla^{l}\nabla^{k}\phi -  \tfrac{3}{4} R_{ikjl} \nabla^{j}\nabla^{i}\phi \nabla^{l}\nabla^{k}\phi -  \tfrac{237}{64} \phi^2 \nabla_{m}R_{ijkl} \nabla^{m}R^{ijkl} \Big) \, .
\end{align}
\endgroup
Notice that $\mathcal{I}_6$ vanishes in flat space after integration by parts.

The tensorial structures in $\mathcal{I}_7$ correspond to Weyl-invariant ordinary anomalies and they are thus solutions of the WZ condition without need of the metric beta-function.
Explicitly, we have
\begingroup
\allowdisplaybreaks[1]
\begin{align}
    \mathcal{I}_7(c_1,h_1,h_2) &= a\, c_1 \Big( -48 h_{1}^{} R_{i}{}^{k} R^{ij} R_{jk} \phi^2 + 4 (9 h_{1}^{} + 8 h_{2}^{}) R_{ij} R^{ij} R \phi^2 \nonumber\\
    &\phantom{=} -  \tfrac{16}{3} (h_{1}^{} + h_{2}^{}) R^3 \phi^2 - 4 (h_{1}^{} + 4 h_{2}^{}) R R_{ijkl} R^{ijkl} \phi^2 - 4 h_{1}^{} R \phi^2 \Box R \nonumber\\
    &\phantom{=} - 4 (5 h_{1}^{} + 8 h_{2}^{}) R^2 \phi \Box \phi - 12 (h_{1}^{} + 8 h_{2}^{}) R_{jklm} R^{jklm} \phi \Box \phi \nonumber\\
    &\phantom{=} - 8 h_{1}^{} R \phi \nabla_{i}\phi \nabla^{i}R + 48 h_{1}^{} R^{jk} \phi \nabla_{i}R_{jk} \nabla^{i}\phi + 24 h_{1}^{} R^{jklm} \phi \nabla_{i}R_{jklm} \nabla^{i}\phi \nonumber\\
    &\phantom{=} - 48 (3 h_{1}^{} + 8 h_{2}^{}) R_{jk} R^{jk} \nabla_{i}\phi \nabla^{i}\phi + 8 (3 h_{1}^{} + 8 h_{2}^{}) R^2 \nabla_{i}\phi \nabla^{i}\phi \nonumber\\
    &\phantom{=} + 24 (3 h_{1}^{} + 8 h_{2}^{}) R_{jklm} R^{jklm} \nabla_{i}\phi \nabla^{i}\phi + 64 h_{1}^{} R^{ij} R \phi \nabla_{j}\nabla_{i}\phi\nonumber\\
    &\phantom{=} - 16 h_{1}^{} R \Box\phi \Box\phi + 48 h_{1}^{} R_{ij} \phi \nabla^{i}R \nabla^{j}\phi + 16 h_{1}^{} R \nabla_{j}\nabla_{i}\phi \nabla^{j}\nabla^{i}\phi \nonumber\\
    &\phantom{=} - 96 h_{1}^{} R^{jk} \phi \nabla^{i}\phi \nabla_{k}R_{ij} - 96 h_{1}^{} R_{i}{}^{k} R^{ij} \phi \nabla_{k}\nabla_{j}\phi + 24 h_{1}^{} R^{ij} \phi^2 \Box R_{ij} \nonumber\\
    &\phantom{=} + 24 (3 h_{1}^{} + 8 h_{2}^{}) R_{ij} R^{ij} \phi \Box\phi + 96 h_{1}^{} R^{ij} \nabla_{j}\nabla_{i}\phi \Box\phi \nonumber\\
    &\phantom{=} - 96 h_{1}^{} R^{ij} \nabla_{k}\nabla_{j}\phi \nabla^{k}\nabla_{i}\phi + 192 h_{1}^{} R_{ijkl} \phi \nabla^{i}\phi \nabla^{l}R^{jk} \nonumber\\
    &\phantom{=} - 48 h_{1}^{} R_{ikjl} \phi^2 \nabla^{l}\nabla^{k}R^{ij} - 96 h_{1}^{} R^{ij} R_{ikjl} \phi \nabla^{l}\nabla^{k}\phi \nonumber\\
    &\phantom{=} - 96 h_{1}^{} R_{ikjl} \nabla^{j}\nabla^{i}\phi \nabla^{l}\nabla^{k}\phi  \Big) \, .
\end{align}
\endgroup
Notice that $\mathcal{I}_7$ vanishes in flat space.
Similar properties hold for $\mathcal{I}_8$ since it is related to $\mathcal{I}_7$ by
\begin{equation}
    \mathcal{I}_8(c_2,h_1,h_2) = \mathcal{I}_7\left(\frac{c_2}{4},h_1,h_2\right) \, .
\end{equation}

The remaining tensorial structures $\mathcal{I}_9,\ldots,\mathcal{I}_{14}$ are parametrised by the free coefficients which appear in~$h^{(2)}_{ij}$.
Each of them solves the WZ condition without need of the metric beta-function and in the field theory they can be eliminated with counterterms.
Thus, they correspond to trivial solutions of the WZ condition.
Explicitly, they read:
\begingroup
\allowdisplaybreaks[1]
\begin{align}
    \mathcal{I}_9(h_{21}) &= a \, h_{21} \Big( -6 R_{i}{}^{k} R^{ij} R_{jk} \phi^2 + 6 R_{ij} R^{ij} R \phi^2 \nonumber\\
    &\phantom{=} -  \tfrac{3}{4} R^3 \phi^2 - 6 R^{ij} R^{kl} R_{ikjl} \phi^2 -  \tfrac{3}{4} R R_{ijkl} R^{ijkl} \phi^2 + 6 R_{i}{}^{m}{}_{k}{}^{n} R^{ijkl} R_{jmln} \phi^2 \nonumber\\
    &\phantom{=} + \tfrac{3}{2} R_{ij}{}^{mn} R^{ijkl} R_{klmn} \phi^2 -  \tfrac{1}{2} R \phi^2 \Box R -  \tfrac{1}{2} R^2 \phi \Box \phi -  \tfrac{3}{2} R_{jklm} R^{jklm} \phi \Box \phi \nonumber\\
    &\phantom{=} -  \tfrac{1}{2} \phi^2 \nabla_{i}R \nabla^{i}R - 2 R \phi \nabla_{i}\phi \nabla^{i}R + 12 R^{jk} \phi \nabla_{i}R_{jk} \nabla^{i}\phi - 6 R^{jklm} \phi \nabla_{i}R_{jklm} \nabla^{i}\phi \nonumber\\
    &\phantom{=} + 3 R_{jk} R^{jk} \nabla_{i}\phi \nabla^{i}\phi -  \tfrac{1}{2} R^2 \nabla_{i}\phi \nabla^{i}\phi -  \tfrac{3}{2} R_{jklm} R^{jklm} \nabla_{i}\phi \nabla^{i}\phi + 3 R^{ij} \phi^2 \Box R_{ij} \nonumber\\
    &\phantom{=} + 3 R_{ij} R^{ij} \phi \Box \phi + 3 \phi^2 \nabla_{k}R_{ij} \nabla^{k}R^{ij} - 6 R_{ikjl} \phi^2 \nabla^{l}\nabla^{k}R^{ij}\nonumber\\
    &\phantom{=} -  \tfrac{3}{2} \phi^2 \nabla_{m}R_{ijkl} \nabla^{m}R^{ijkl} \Big)
\end{align}
\endgroup

\begingroup
\allowdisplaybreaks[1]
\begin{align}
    \mathcal{I}_{10}(h_{22}) &= a \, h_{22} \Big( 3 \nabla^{j}\Box\phi \Box \nabla_{j}\phi - 3 \nabla_{k}\nabla_{j}\nabla_{i}\phi \nabla^{k}\nabla^{j}\nabla^{i}\phi + 3 R_{i}{}^{k} R^{ij} R_{jk} \phi^2 \nonumber\\
    &\phantom{=} -  \tfrac{3}{2} R_{ij} R^{ij} R \phi^2 + \tfrac{3}{16} R^3 \phi^2 + \tfrac{3}{16} R R_{ijkl} R^{ijkl} \phi^2 -  \tfrac{3}{2} R_{i}{}^{m}{}_{k}{}^{n} R^{ijkl} R_{jmln} \phi^2 \nonumber\\
    &\phantom{=} -  \tfrac{3}{8} R_{ij}{}^{mn} R^{ijkl} R_{klmn} \phi^2 + \tfrac{1}{2} R \phi^2 \Box R + \tfrac{3}{4} R^2 \phi \Box \phi + \tfrac{3}{4} R_{jklm} R^{jklm} \phi \Box \phi \nonumber\\
    &\phantom{=} + \tfrac{5}{8} \phi^2 \nabla_{i}R \nabla^{i}R + \tfrac{3}{2} R \phi \nabla_{i}\phi \nabla^{i}R - 9 R^{jk} \phi \nabla_{i}R_{jk} \nabla^{i}\phi + \tfrac{9}{4} R^{jklm} \phi \nabla_{i}R_{jklm} \nabla^{i}\phi \nonumber\\
    &\phantom{=} -  \tfrac{3}{2} R_{jk} R^{jk} \nabla_{i}\phi \nabla^{i}\phi + \tfrac{3}{8} R^2 \nabla_{i}\phi \nabla^{i}\phi + \tfrac{3}{8} R_{jklm} R^{jklm} \nabla_{i}\phi \nabla^{i}\phi + R^{ij} \phi^2 \nabla_{j}\nabla_{i}R \nonumber\\
    &\phantom{=} - 2 R^{ij} R \phi \nabla_{j}\nabla_{i}\phi + \phi \Box R \Box \phi + R \Box \phi \Box \phi + \tfrac{1}{2} \nabla_{i}\phi \nabla^{i}R \Box \phi + \tfrac{3}{2} \phi \nabla^{i}R \Box \nabla_{i}\phi \nonumber\\
    &\phantom{=} -  R_{ij} \phi \nabla^{i}R \nabla^{j}\phi + 3 R_{i}{}^{k} R_{jk} \nabla^{i}\phi \nabla^{j}\phi -  \tfrac{3}{2} R_{ij} R \nabla^{i}\phi \nabla^{j}\phi + 3 R^{kl} R_{ikjl} \nabla^{i}\phi \nabla^{j}\phi \nonumber\\
    &\phantom{=} -  \tfrac{1}{2} \nabla^{i}R \nabla_{j}\nabla_{i}\phi \nabla^{j}\phi + 2 \phi \nabla_{j}\nabla_{i}\phi \nabla^{j}\nabla^{i}R -  R \nabla_{j}\nabla_{i}\phi \nabla^{j}\nabla^{i}\phi \nonumber\\
    &\phantom{=} + 3 R^{jk} \phi \nabla^{i}\phi \nabla_{k}R_{ij} + 6 R_{i}{}^{k} R^{ij} \phi \nabla_{k}\nabla_{j}\phi - 3 R^{jk} \nabla^{i}\phi \nabla_{k}\nabla_{j}\nabla_{i}\phi \nonumber\\
    &\phantom{=} - 3 R^{ij} \phi^2 \Box R_{ij}  - 6 \phi \nabla^{j}\nabla^{i}\phi \Box R_{ij} - 3 R_{ij} R^{ij} \phi \Box \phi - 6 R^{ij} \nabla_{j}\nabla_{i}\phi \Box \phi \nonumber\\
    &\phantom{=} + 6 R_{i}{}^{j} \nabla^{i}\phi \Box \nabla_{j}\phi  + 3 \phi^2 \nabla_{j}R_{ik} \nabla^{k}R^{ij} -  \tfrac{9}{2} \phi^2 \nabla_{k}R_{ij} \nabla^{k}R^{ij} \nonumber\\
    &\phantom{=} - 3 \phi \nabla_{k}\nabla_{j}\nabla_{i}\phi \nabla^{k}R^{ij}  + 3 R^{ij} \nabla_{k}\nabla_{j}\phi \nabla^{k}\nabla_{i}\phi - 3 \nabla_{i}R_{jk} \nabla^{i}\phi \nabla^{k}\nabla^{j}\phi \nonumber\\
    &\phantom{=} + 3 \nabla^{i}\phi \nabla_{k}R_{ij} \nabla^{k}\nabla^{j}\phi  - 3 R_{ijkl} \phi \nabla^{i}\phi \nabla^{l}R^{jk} + 9 R_{ikjl} \nabla^{j}\nabla^{i}\phi \nabla^{l}\nabla^{k}\phi \nonumber\\
    &\phantom{=} + \tfrac{3}{8} \phi^2 \nabla_{m}R_{ijkl} \nabla^{m}R^{ijkl} \Big)
\end{align}
\endgroup

\begingroup
\allowdisplaybreaks[1]
\begin{align}
    \mathcal{I}_{11}(h_{23}) &= a \, h_{23} \Big( \tfrac{3}{2} \nabla^{j}\Box \phi \Box\nabla_{j}\phi -  \tfrac{3}{2} \nabla_{k}\nabla_{j}\nabla_{i}\phi \nabla^{k}\nabla^{j}\nabla^{i}\phi - \tfrac{3}{2} R_{i}{}^{k} R^{ij} R_{jk} \phi^2 \nonumber\\
    &\phantom{=} + \tfrac{9}{4} R_{ij} R^{ij} R \phi^2 -  \tfrac{9}{32} R^3 \phi^2 - 3 R^{ij} R^{kl} R_{ikjl} \phi^2 -  \tfrac{9}{32} R R_{ijkl} R^{ijkl} \phi^2 \nonumber\\
    &\phantom{=} + \tfrac{9}{4} R_{i}{}^{m}{}_{k}{}^{n} R^{ijkl} R_{jmln} \phi^2 + \tfrac{9}{16} R_{ij}{}^{mn} R^{ijkl} R_{klmn} \phi^2 + \tfrac{1}{8} R^2 \phi \Box \phi \nonumber\\
    &\phantom{=} -  \tfrac{3}{8} R_{jklm} R^{jklm} \phi \Box \phi + \tfrac{1}{16} \phi^2 \nabla_{i}R \nabla^{i}R -  \tfrac{1}{4} R \phi \nabla_{i}\phi \nabla^{i}R + \tfrac{3}{2} R^{jk} \phi \nabla_{i}R_{jk} \nabla^{i}\phi \nonumber\\
    &\phantom{=} -  \tfrac{15}{8} R^{jklm} \phi \nabla_{i}R_{jklm} \nabla^{i}\phi + \tfrac{3}{4} R_{jk} R^{jk} \nabla_{i}\phi \nabla^{i}\phi -  \tfrac{1}{16} R^2 \nabla_{i}\phi \nabla^{i}\phi \nonumber\\
    &\phantom{=} -  \tfrac{9}{16} R_{jklm} R^{jklm} \nabla_{i}\phi \nabla^{i}\phi + \tfrac{1}{2} R^{ij} \phi^2 \nabla_{j}\nabla_{i}R -  R^{ij} R \phi \nabla_{j}\nabla_{i}\phi + \tfrac{1}{2} \phi \Box R \Box \phi \nonumber\\
    &\phantom{=} + \tfrac{1}{2} R \Box \phi \Box \phi + \tfrac{1}{4} \nabla_{i}\phi \nabla^{i}R \Box \phi + \tfrac{3}{4} \phi \nabla^{i}R \Box \nabla_{i}\phi -  \tfrac{1}{2} R_{ij} \phi \nabla^{i}R \nabla^{j}\phi \nonumber\\
    &\phantom{=} + \tfrac{3}{2} R_{i}{}^{k} R_{jk} \nabla^{i}\phi \nabla^{j}\phi -  \tfrac{3}{4} R_{ij} R \nabla^{i}\phi \nabla^{j}\phi + \tfrac{3}{2} R^{kl} R_{ikjl} \nabla^{i}\phi \nabla^{j}\phi \nonumber\\
    &\phantom{=} -  \tfrac{1}{4} \nabla^{i}R \nabla_{j}\nabla_{i}\phi \nabla^{j}\phi + \phi \nabla_{j}\nabla_{i}\phi \nabla^{j}\nabla^{i}R -  \tfrac{1}{2} R \nabla_{j}\nabla_{i}\phi \nabla^{j}\nabla^{i}\phi \nonumber\\
    &\phantom{=} + \tfrac{3}{2} R^{jk} \phi \nabla^{i}\phi \nabla_{k}R_{ij} + 3 R_{i}{}^{k} R^{ij} \phi \nabla_{k}\nabla_{j}\phi -  \tfrac{3}{2} R^{jk} \nabla^{i}\phi \nabla_{k}\nabla_{j}\nabla_{i}\phi \nonumber\\
    &\phantom{=} - 3 \phi \nabla^{j}\nabla^{i}\phi \Box R_{ij} - 3 R^{ij} \nabla_{j}\nabla_{i}\phi \Box \phi + 3 R_{i}{}^{j} \nabla^{i}\phi \Box \nabla_{j}\phi + \tfrac{3}{2} \phi^2 \nabla_{j}R_{ik} \nabla^{k}R^{ij} \nonumber\\
    &\phantom{=} -  \tfrac{3}{4} \phi^2 \nabla_{k}R_{ij} \nabla^{k}R^{ij} -  \tfrac{3}{2} \phi \nabla_{k}\nabla_{j}\nabla_{i}\phi \nabla^{k}R^{ij} + \tfrac{3}{2} R^{ij} \nabla_{k}\nabla_{j}\phi \nabla^{k}\nabla_{i}\phi \nonumber\\
    &\phantom{=} -  \tfrac{3}{2} \nabla_{i}R_{jk} \nabla^{i}\phi \nabla^{k}\nabla^{j}\phi + \tfrac{3}{2} \nabla^{i}\phi \nabla_{k}R_{ij} \nabla^{k}\nabla^{j}\phi -  \tfrac{3}{2} R_{ijkl} \phi \nabla^{i}\phi \nabla^{l}R^{jk} \nonumber\\
    &\phantom{=} - 3 R_{ikjl} \phi^2 \nabla^{l}\nabla^{k}R^{ij} + \tfrac{9}{2} R_{ikjl} \nabla^{j}\nabla^{i}\phi \nabla^{l}\nabla^{k}\phi -  \tfrac{9}{16} \phi^2 \nabla_{m}R_{ijkl} \nabla^{m}R^{ijkl} \Big)
\end{align}
\endgroup

\begingroup
\allowdisplaybreaks[1]
\begin{align}
    \mathcal{I}_{12}(h_{24}) &= a \, h_{24} \Big( 3 \nabla^{j}\Box\phi \Box\nabla_{j}\phi - 3 \nabla_{k}\nabla_{j}\nabla_{i}\phi \nabla^{k}\nabla^{j}\nabla^{i}\phi -6 R_{i}{}^{k} R^{ij} R_{jk} \phi^2 \nonumber\\
    &\phantom{=} + \tfrac{15}{2} R_{ij} R^{ij} R \phi^2 -  \tfrac{15}{16} R^3 \phi^2 - 9 R^{ij} R^{kl} R_{ikjl} \phi^2 -  \tfrac{15}{16} R R_{ijkl} R^{ijkl} \phi^2 \nonumber\\
    &\phantom{=} + \tfrac{15}{2} R_{i}{}^{m}{}_{k}{}^{n} R^{ijkl} R_{jmln} \phi^2 + \tfrac{15}{8} R_{ij}{}^{mn} R^{ijkl} R_{klmn} \phi^2 -  \tfrac{1}{4} R \phi^2 \Box R + \tfrac{1}{2} R^2 \phi \Box \phi \nonumber\\
    &\phantom{=} -  \tfrac{3}{2} R_{jklm} R^{jklm} \phi \Box \phi -  \tfrac{1}{8} \phi^2 \nabla_{i}R \nabla^{i}R -  \tfrac{3}{2} R \phi \nabla_{i}\phi \nabla^{i}R + 9 R^{jk} \phi \nabla_{i}R_{jk} \nabla^{i}\phi \nonumber\\
    &\phantom{=} -  \tfrac{27}{4} R^{jklm} \phi \nabla_{i}R_{jklm} \nabla^{i}\phi + \tfrac{3}{2} R_{jk} R^{jk} \nabla_{i}\phi \nabla^{i}\phi + \tfrac{1}{8} R^2 \nabla_{i}\phi \nabla^{i}\phi \nonumber\\
    &\phantom{=} -  \tfrac{15}{8} R_{jklm} R^{jklm} \nabla_{i}\phi \nabla^{i}\phi + R^{ij} \phi^2 \nabla_{j}\nabla_{i}R - 4 R^{ij} R \phi \nabla_{j}\nabla_{i}\phi -  \tfrac{1}{2} \nabla_{i}\phi \nabla^{i}\phi \Box R \nonumber\\
    &\phantom{=} + \tfrac{1}{2} \phi \Box R \Box \phi + R \Box\phi \Box\phi + \tfrac{1}{2} \nabla_{i}\phi \nabla^{i}R \Box \phi + \tfrac{3}{2} \phi \nabla^{i}R \Box\nabla_{i}\phi -  R_{ij} \phi \nabla^{i}R \nabla^{j}\phi \nonumber\\
    &\phantom{=} + 3 R_{i}{}^{k} R_{jk} \nabla^{i}\phi \nabla^{j}\phi -  \tfrac{7}{2} R_{ij} R \nabla^{i}\phi \nabla^{j}\phi + 9 R^{kl} R_{ikjl} \nabla^{i}\phi \nabla^{j}\phi \nonumber\\
    &\phantom{=} -  \nabla^{i}\phi \nabla_{j}\nabla_{i}R \nabla^{j}\phi -  \tfrac{1}{2} \nabla^{i}R \nabla_{j}\nabla_{i}\phi \nabla^{j}\phi + \phi \nabla_{j}\nabla_{i}\phi \nabla^{j}\nabla^{i}R \nonumber\\
    &\phantom{=} -  R \nabla_{j}\nabla_{i}\phi \nabla^{j}\nabla^{i}\phi + 3 R^{jk} \phi \nabla^{i}\phi \nabla_{k}R_{ij} + 6 R_{i}{}^{k} R^{ij} \phi \nabla_{k}\nabla_{j}\phi \nonumber\\
    &\phantom{=} - 3 R^{jk} \nabla^{i}\phi \nabla_{k}\nabla_{j}\nabla_{i}\phi + \tfrac{3}{2} R^{ij} \phi^2 \Box R_{ij} + 3 \nabla^{i}\phi \nabla^{j}\phi \Box R_{ij} - 3 \phi \nabla^{j}\nabla^{i}\phi \Box R_{ij} \nonumber\\
    &\phantom{=} - 6 R^{ij} \nabla_{j}\nabla_{i}\phi \Box \phi + 6 R_{i}{}^{j} \nabla^{i}\phi \Box \nabla_{j}\phi + 3 \phi^2 \nabla_{j}R_{ik} \nabla^{k}R^{ij} \nonumber\\
    &\phantom{=} - 3 \phi \nabla_{k}\nabla_{j}\nabla_{i}\phi \nabla^{k}R^{ij} + 3 R^{ij} \nabla_{k}\nabla_{j}\phi \nabla^{k}\nabla_{i}\phi - 3 \nabla_{i}R_{jk} \nabla^{i}\phi \nabla^{k}\nabla^{j}\phi \nonumber\\
    &\phantom{=} + 3 \nabla^{i}\phi \nabla_{k}R_{ij} \nabla^{k}\nabla^{j}\phi - 3 R_{ijkl} \phi \nabla^{i}\phi \nabla^{l}R^{jk} - 9 R_{ikjl} \phi^2 \nabla^{l}\nabla^{k}R^{ij} \nonumber\\
    &\phantom{=} + 6 R^{ij} R_{ikjl} \phi \nabla^{l}\nabla^{k}\phi + 9 R_{ikjl} \nabla^{j}\nabla^{i}\phi \nabla^{l}\nabla^{k}\phi -  \tfrac{15}{8} \phi^2 \nabla_{m}R_{ijkl} \nabla^{m}R^{ijkl} \Big)
\end{align}
\endgroup

\begingroup
\allowdisplaybreaks[1]
\begin{align}
    \mathcal{I}_{13}(h_{25}) &= a \, h_{25} \Big( \tfrac{1}{2} \nabla^{j}\Box \phi \Box \nabla_{j}\phi - 4 \nabla^{j}\nabla^{i}\phi \Box \nabla_{j}\nabla_{i}\phi + \Box\phi \Box^2\phi \nonumber\\
    &\phantom{=} -  \tfrac{7}{2} \nabla_{k}\nabla_{j}\nabla_{i}\phi \nabla^{k}\nabla^{j}\nabla^{i}\phi - \tfrac{5}{2} R_{i}{}^{k} R^{ij} R_{jk} \phi^2 + \tfrac{9}{4} R_{ij} R^{ij} R \phi^2 -  \tfrac{9}{32} R^3 \phi^2 \nonumber\\
    &\phantom{=} - 2 R^{ij} R^{kl} R_{ikjl} \phi^2 -  \tfrac{9}{32} R R_{ijkl} R^{ijkl} \phi^2 + \tfrac{9}{4} R_{i}{}^{m}{}_{k}{}^{n} R^{ijkl} R_{jmln} \phi^2 \nonumber\\
    &\phantom{=} + \tfrac{9}{16} R_{ij}{}^{mn} R^{ijkl} R_{klmn} \phi^2 -  \tfrac{3}{4} R_{jklm} R^{jklm} \phi \Box\phi -  \tfrac{1}{16} \phi^2 \nabla_{i}R \nabla^{i}R \nonumber\\
    &\phantom{=} + 2 R^{jk} \phi \nabla_{i}R_{jk} \nabla^{i}\phi -  \tfrac{21}{8} R^{jklm} \phi \nabla_{i}R_{jklm} \nabla^{i}\phi + \tfrac{3}{4} R_{jk} R^{jk} \nabla_{i}\phi \nabla^{i}\phi \nonumber\\
    &\phantom{=} -  \tfrac{1}{16} R^2 \nabla_{i}\phi \nabla^{i}\phi -  \tfrac{13}{16} R_{jklm} R^{jklm} \nabla_{i}\phi \nabla^{i}\phi -  \tfrac{1}{4} R^{ij} \phi^2 \nabla_{j}\nabla_{i}R + \tfrac{1}{4} \phi \Box R \Box \phi \nonumber\\
    &\phantom{=} + \tfrac{1}{2} R \Box\phi \Box \phi + \tfrac{3}{4} \nabla_{i}\phi \nabla^{i}R \Box \phi + \tfrac{3}{4} \phi \nabla^{i}R \Box \nabla_{i}\phi + R \nabla^{i}\phi \Box \nabla_{i}\phi + \tfrac{1}{2} R \phi \Box^2 \phi \nonumber\\
    &\phantom{=} -  R_{ij} \phi \nabla^{i}R \nabla^{j}\phi + \tfrac{1}{2} R_{i}{}^{k} R_{jk} \nabla^{i}\phi \nabla^{j}\phi -  \tfrac{3}{4} R_{ij} R \nabla^{i}\phi \nabla^{j}\phi + \tfrac{7}{2} R^{kl} R_{ikjl} \nabla^{i}\phi \nabla^{j}\phi \nonumber\\
    &\phantom{=} + \tfrac{1}{4} \nabla^{i}R \nabla_{j}\nabla_{i}\phi \nabla^{j}\phi -  \tfrac{1}{2} \phi \nabla_{j}\nabla_{i}\phi \nabla^{j}\nabla^{i}R -  \tfrac{1}{2} R^{jk} \phi \nabla^{i}\phi \nabla_{k}R_{ij} \nonumber\\
    &\phantom{=} -  R_{i}{}^{k} R^{ij} \phi \nabla_{k}\nabla_{j}\phi -  \tfrac{7}{2} R^{jk} \nabla^{i}\phi \nabla_{k}\nabla_{j}\nabla_{i}\phi + \tfrac{1}{2} R^{ij} \phi^2 \Box R_{ij} -  \tfrac{1}{2} \phi \nabla^{j}\nabla^{i}\phi \Box R_{ij} \nonumber\\
    &\phantom{=} + \tfrac{1}{2} R_{ij} R^{ij} \phi \Box \phi - 2 R^{ij} \nabla_{j}\nabla_{i}\phi \Box \phi - 2 R_{i}{}^{j} \nabla^{i}\phi \Box \nabla_{j}\phi - 2 R^{ij} \phi \Box\nabla_{j}\nabla_{i}\phi \nonumber\\
    &\phantom{=} -  \phi^2 \nabla_{j}R_{ik} \nabla^{k}R^{ij} + \tfrac{5}{4} \phi^2 \nabla_{k}R_{ij} \nabla^{k}R^{ij} -  \tfrac{7}{2} \phi \nabla_{k}\nabla_{j}\nabla_{i}\phi \nabla^{k}R^{ij} \nonumber\\
    &\phantom{=} + \tfrac{1}{2} R^{ij} \nabla_{k}\nabla_{j}\phi \nabla^{k}\nabla_{i}\phi -  \tfrac{5}{2} \nabla_{i}R_{jk} \nabla^{i}\phi \nabla^{k}\nabla^{j}\phi -  \tfrac{3}{2} \nabla^{i}\phi \nabla_{k}R_{ij} \nabla^{k}\nabla^{j}\phi \nonumber\\
    &\phantom{=} -  \tfrac{1}{2} R_{ijkl} \phi \nabla^{i}\phi \nabla^{l}R^{jk} -  \tfrac{3}{2} R_{ikjl} \phi^2 \nabla^{l}\nabla^{k}R^{ij} + R^{ij} R_{ikjl} \phi \nabla^{l}\nabla^{k}\phi \nonumber\\
    &\phantom{=} -  \tfrac{1}{2} R_{ikjl} \nabla^{j}\nabla^{i}\phi \nabla^{l}\nabla^{k}\phi -  \tfrac{9}{16} \phi^2 \nabla_{m}R_{ijkl} \nabla^{m}R^{ijkl} \Big)
\end{align}
\endgroup

\begingroup
\allowdisplaybreaks[1]
\begin{align}
    \mathcal{I}_{14}(h_{26}) &= a \, h_{26} \Big( 3 \nabla^{j}\Box\phi \Box\nabla_{j}\phi - 12 \nabla^{j}\nabla^{i}\phi \Box\nabla_{j}\nabla_{i}\phi + 3 \Box\phi \Box^2\phi \nonumber\\
    &\phantom{=} - 12 \nabla_{k}\nabla_{j}\nabla_{i}\phi \nabla^{k}\nabla^{j}\nabla^{i}\phi -6 R_{i}{}^{k} R^{ij} R_{jk} \phi^2 + 6 R_{ij} R^{ij} R \phi^2 -  \tfrac{3}{4} R^3 \phi^2 \nonumber\\
    &\phantom{=} - 6 R^{ij} R^{kl} R_{ikjl} \phi^2 -  \tfrac{3}{4} R R_{ijkl} R^{ijkl} \phi^2 + 6 R_{i}{}^{m}{}_{k}{}^{n} R^{ijkl} R_{jmln} \phi^2 \nonumber\\
    &\phantom{=} + \tfrac{3}{2} R_{ij}{}^{mn} R^{ijkl} R_{klmn} \phi^2 + \tfrac{1}{4} R \phi^2 \Box R + \tfrac{1}{4} R^2 \phi \Box \phi -  \tfrac{3}{2} R_{jklm} R^{jklm} \phi \Box \phi \nonumber\\
    &\phantom{=} + \tfrac{1}{4} \phi^2 \nabla_{i}R \nabla^{i}R + R \phi \nabla_{i}\phi \nabla^{i}R - 6 R^{jklm} \phi \nabla_{i}R_{jklm} \nabla^{i}\phi + \tfrac{1}{4} R^2 \nabla_{i}\phi \nabla^{i}\phi \nonumber\\
    &\phantom{=} -  \tfrac{3}{2} R_{jklm} R^{jklm} \nabla_{i}\phi \nabla^{i}\phi + \tfrac{3}{2} \phi \Box R \Box \phi + \tfrac{3}{2} R \Box \phi \Box \phi + 3 \nabla_{i}\phi \nabla^{i}R \Box \phi \nonumber\\
    &\phantom{=} + 3 \phi \nabla^{i}R \Box \nabla_{i}\phi + 3 R \nabla^{i}\phi \Box \nabla_{i}\phi + \tfrac{3}{2} R \phi \Box^2\phi - 3 R_{ij} \phi \nabla^{i}R \nabla^{j}\phi \nonumber\\
    &\phantom{=} + 3 R_{i}{}^{k} R_{jk} \nabla^{i}\phi \nabla^{j}\phi - 3 R_{ij} R \nabla^{i}\phi \nabla^{j}\phi + 12 R^{kl} R_{ikjl} \nabla^{i}\phi \nabla^{j}\phi \nonumber\\
    &\phantom{=} - 12 R^{jk} \nabla^{i}\phi \nabla_{k}\nabla_{j}\nabla_{i}\phi - 6 \phi \nabla^{j}\nabla^{i}\phi \Box R_{ij} - 6 R^{ij} \nabla_{j}\nabla_{i}\phi \Box\phi \nonumber\\
    &\phantom{=} - 3 R_{i}{}^{j} \nabla^{i}\phi \Box \nabla_{j}\phi - 6 R^{ij} \phi \Box \nabla_{j}\nabla_{i}\phi - 12 \phi \nabla_{k}\nabla_{j}\nabla_{i}\phi \nabla^{k}R^{ij} \nonumber\\
    &\phantom{=} - 12 \nabla_{i}R_{jk} \nabla^{i}\phi \nabla^{k}\nabla^{j}\phi - 6 R_{ikjl} \phi^2 \nabla^{l}\nabla^{k}R^{ij} -  \tfrac{3}{2} \phi^2 \nabla_{m}R_{ijkl} \nabla^{m}R^{ijkl}  \Big)
\end{align}
\endgroup
Notice that $\mathcal{I}_9$ vanishes in flat space and similarly  $\mathcal{I}_{10},\ldots,\mathcal{I}_{14}$ after integration by parts.


\newpage

\phantomsection
\addcontentsline{toc}{section}{References}
\bibliography{biblio}

\end{document}